\documentclass[11pt,letterpaper]{article}

\usepackage[T1]{fontenc}
\usepackage[english]{babel}
\usepackage{times}

\usepackage[letterpaper,margin=1in]{geometry}

\usepackage{amsmath,amstext,amsbsy,amsopn,amsthm,amsfonts,amssymb}
\usepackage{mathtools}
\usepackage{url}
\usepackage{tabularx,booktabs}
\newcolumntype{C}{>{\centering\arraybackslash}X}
\usepackage{booktabs,array,colortbl}
 \usepackage[]{graphicx}
\usepackage{rotate}
\usepackage{wrapfig}
\usepackage{caption}
\usepackage{subfigure}
\usepackage{lastpage}

\usepackage[small]{titlesec}
\usepackage{paralist}

\usepackage{verbatim} 
\usepackage[usenames,dvipsnames,svgnames]{xcolor}
\usepackage{multirow}
\usepackage{algorithmic}
\usepackage{algorithm}

\usepackage{fancyhdr}
\providecommand{\keywords}[1]{\textbf{\textit{Keywords---}} #1}

\renewenvironment{thebibliography}[1]{%
	\begin{oldthebibliography}{#1}%
		\setlength{\parskip}{0.0cm}%
		\setlength{\itemsep}{0.0cm}%
	}%
	{%
	\end{oldthebibliography}%
}

\begin{document}


\title{\textbf{cSeiz}: An Edge-Device for Accurate Seizure Detection and Control for Smart Healthcare}

\newcommand{\matlab}[0]{MATLAB\textsuperscript{\textregistered}{\hspace{0.03cm}}}
\newcommand{\simulink}[0]{Simulink\textsuperscript{\textregistered}{\hspace{0.03cm}}}
\newcommand{\simscape}[0]{Simscape\textsuperscript{\textregistered}{\hspace{0.03cm}}}
\newcommand{\etal}[0]{\emph{et~al.}}

\maketitle

\author{
	\begin{center}
\begin{tabular}{ccc}
Md Abu Sayeed  &			Saraju P. Mohanty &  Elias Kougianos \\
Computer Science and Engineering &		Computer Science and Engineering  &		Electrical Engineering \\
University of North Texas, USA.	&		University of North Texas, USA.  &		University of North Texas, USA. \\
\texttt{mdsayeed@my.unt.edu} 	&		\texttt{saraju.mohanty@unt.edu} &  \texttt{elias.kougianos@unt.edu} \\\\
\end{tabular}
	\end{center}
}

\cfoot{Page -- \thepage-of-\pageref{LastPage}}

\begin{abstract}
Epilepsy is one of the most common neurological disorders affecting up to 1\% of the world's population and approximately 2.5 million people in the United States. Seizures in more than 30\% of epilepsy patients are refractory to anti-epileptic drugs. An important biomedical research effort is focused on the development of an energy efficient implantable device for the real-time control of seizures. In this paper we propose an Internet of Medical Things (IoMT) based automated seizure detection and drug delivery system (DDS) for the control of seizures. The proposed system will detect seizures and inject a fast acting anti-convulsant drug at the onset to suppress seizure progression. The drug injection is performed in two stages. Initially, the seizure detector detects the seizure from the electroencephalography (EEG) signal using a hyper-synchronous signal detection circuit and a signal rejection algorithm (SRA). In the second stage, the drug is released in the seizure onset area upon seizure detection. The design was validated using a system-level simulation and consumer electronics proof of concept. The proposed seizure detector reports a sensitivity of 96.9\% and specificity of 97.5\%. The use of minimal circuitry leads to a considerable reduction of power consumption compared to previous approaches. The proposed approach can be generalized to other sensor modalities and the use of both wearable and implantable solutions, or a combination of the two. 
\end{abstract}

\keywords{
Smart Healthcare, IoMT, Wearable Medical Device (WDM), Implantable Medical Device (IMD), EEG, Seizure, Epilepsy, Valveless Micro-pump, Drug Delivery, Closed Loop Control, Energy Efficient Systems, Low Latency Systems
}


\section{Introduction}


Smart health care is becoming very important due to the combined pressures of an increasing population, an increasing demand for excellent care and limited resources. Smart health achieved through wearables has been focused on general wellness, but is now starting to encompass the management of acute disorders. One example of such an effort is the use of smart health care for the automated real-time control of seizures. Epilepsy is a neurological disorder marked by spontaneous recurrent seizures. A seizure is the manifestation of an abnormal hyper-synchronous disturbance of a population of neurons \cite{Mormann_Brain_2007}, which may manifest as sensory disturbance, loss of awareness, or convulsions.  

There are multiple lines of treatment for epilepsy. Anti-convulsant drugs are used as a first line to control seizures, but more than 30\% of epilepsy patients remain refractory to medication.
For patients who are refractory, uncontrolled seizures have a devastating impact on the patient's quality of life. Epilepsy surgery is suitable for some patients with medically refractory seizures, but not if the patient has multi-focal seizures or if the seizure onset area is located in the eloquent cortex \cite{Spencer_Lancet_2008}. Other possibilities include modification of diet, which may be effective in some children. Wearable and implantable devices constitute an important fourth line of treatment, and one whose use is growing. In this approach wearable or fully implantable devices are used for automated monitoring, warning and suppression of seizures. Early warning can enable a patient to take protective action when necessary \cite{Epilepsy_Book_Osorio_2011}, \cite{Gluckman_JCN_2015}. Automated, closed loop therapy has shown efficacy in managing epilepsy. Responsive neural stimulation (RNS), for example, which is approved by the Food and Drug Administration (FDA) for use in the USA has been shown to reduce the number of seizures experienced by a patient. 

\begin{figure}[htbp]
	\centering
	\includegraphics[width=0.65\textwidth]{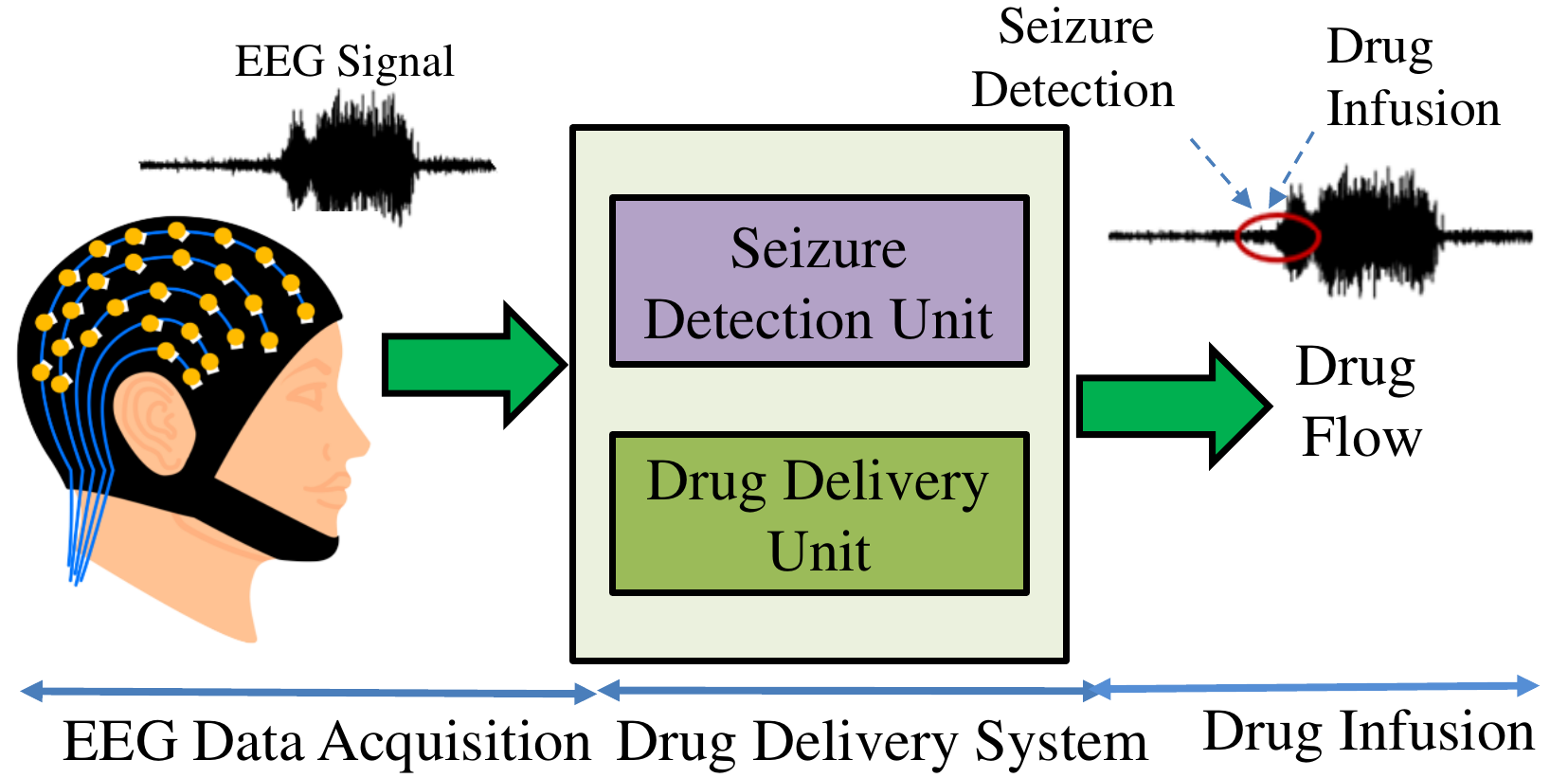}
	\caption{Seizure Detection and Drug Delivery System Based on an EEG.}
	\label{FIG:model_drug_delivery_system}
\end{figure}

Electroencephalography (EEG) signals serve as input into the seizure detector and drug delivery system. Upon detection of a seizure, a fast-acting drug is delivered to terminate the seizure. The EEG can be collected in a number of ways, including through the use of a cap, headband, and invasive or subcutaneous sensors. Also possible is the use of other wearable sensors to monitor galvanic skin response and heart rate to detect changes within the autonomic nervous system reflective of seizure onset. In a wearable solution a fast acting drug can be delivered as a nasal spray, while in an implantable system a fast acting drug can be infused into the seizure onset area. 

In this paper we propose a seizure control system composed of an automated seizure detector and drug delivery system. The proposed seizure detector offers reduced design complexity, power consumption, and sensitivity to noise. In this approach, neural signals are monitored continuously. A detection circuit analyzes the input signal and detects candidate seizure activity (hyper-synchronization). A separate algorithm is used to analyze the hyper-synchronous pulses. If the pulses exceed a threshold, then a seizure is declared.
\textcolor{black}{An open loop system provides continuous release of drugs at regular time intervals, while a closed loop system injects the drug only once, when a seizure occurs}. In this paper, a closed loop drug delivery system is proposed in which an anti-convulsant drug is infused directly into the seizure onset area upon seizure detection. An illustration of the proposed system is shown in Fig. \ref{FIG:model_drug_delivery_system}. The proposed system consists of two parts: the seizure detection unit and the drug delivery unit. The seizure detection unit consists of a filter, an amplifier, a voltage level detector, and a signal rejection algorithm. The drug delivery unit consists of an actuator element, valveless micro-pump and reservoir. 

The remainder of the manuscript is organized as follows: Section \ref{Sec:Novel_Contributions} highlights the novel contributions. Existing research on seizure detection is summarized in Section \ref{Sec:Prior_Research}. Section \ref{Sec:System_Overview} illustrates the architectural overview and design of the proposed system. The proposed seizure detector and drug delivery system is illustrated in Sections \ref{Sec:Proposed_Detector} and  \ref{Sec:Proposed_Delivery-System}, respectively. The implementation of the proposed design is discussed in Section \ref{Sec:Proposed_Modeling}. The simulation results are shown in Section \ref{Sec:Experimental_Results} and conclusions are presented in Section \ref{Sec:Conclusions}.


\section{Novel Contributions of The Current Paper}
\label{Sec:Novel_Contributions}


\begin{enumerate}
	
\item	
The proposed signal rejection algorithm (SRA) analyzes the data and then accurately removes unwanted bursts of pulse and high frequency samples. The algorithm continues until the hyper-synchronous  pulses cross a threshold point, which helps to reduce false detections and provides better detection accuracy. The use of limited circuitry in the proposed seizure detector leads to considerable reduction in power consumption.
 
\item 
Micro-pumps are useful tools for drug delivery as they maintain accurate flow control and low power consumption. The proposed micro-pump does not have a valve that opens and closes during the pumping cycle which makes it simple and less expensive compared to commercially available micro-pumps. 

\item 
Existing devices do not include the IoMT for remote connectivity. The integration of the proposed system with the IoMT provides the advantages of remote healthcare monitoring and consultation. The patient's EEG data are continuously stored and analyzed on the cloud  and a notification is sent to the physician if a seizure occurs. Doctors can prescribe the required drug based on the healthcare report from the cloud.

\item 
The proposed system provides a considerable reduction in power consumption, and increased detection accuracy making it suitable for use as a wearable or implantable device for seizure detection and control through drug delivery.

\end{enumerate}

\section{Related Prior Research}
\label{Sec:Prior_Research}

Internet-of-Medical-Things (IoMT) based smart healthcare can have both Wearable Medical Devices (WMDs) and Implantable Medical Devices (IMDs) which can be collectively called Implantable and Wearable Medical Devices (IWMD) \cite{Zhang_JPROC_2014-Aug, Sundaravadivel_IEEE-CEM_2018}. The IoMT devices are connected to the network by various methods, wired or wireless which gives them communication capabilities and in many cases makes
them mobile \cite{Sundaravadivel_IEEE-CEM_2018, Yanambaka_TCE_2019-Aug_PMsec}. Most of the elements of IoMT are essentially consumer electronics (CE) devices \cite{Mohanty_ISCT-2019_Keynote-Talk_CT-Smart-City}.

Table \ref{TABLE:existing_smart_healthcare} shows relevant existing research and development in consumer electronics and illustrates their contributions to smart healthcare. An IoMT based smart healthcare system has been proposed in \cite{Ullah_ICISE_2016}, which collects patient's medical data including blood pressure and heart rate, and sends them to a physician through the cloud, providing remote healthcare service. An architecture of remote electrocardiogram (ECG) monitoring with a wireless sensor network has been proposed in \cite{Dey_IEEE-TCE_2017, Ivanov_IEEE-TCE_2012, Spinsante_IEEE-TCE_2012, Raj_IEEE-TCE_2018} in which portable sensors transmit data to an ECG server which are then sent to hospitals and physicians for evaluation. An extensive survey has been conducted in \cite{Sundaravadivel_IEEE-CEM_2018} which reviews the concepts, applications, current research trends, challenges, opportunities and significance of  IoMT in smart health care.  U.S regulators have recently approved the first medical grade smart watch, a novel consumer electronics product for neurological health, which measures abnormal  activity during an epileptic seizure and sends alerts to a physician for proper action \cite{Dolgin_IEEE-spectrum_2018}. While this product has shown promising results for the detection of generalized clonic-tonic seizures, significant research remains to be conducted for the detection of partial seizures and development of efficient drug delivery systems. The proposed system advances consumer electronics by bringing seizure detection and control to the smart health care system.

\begin{table*}[htbp]
	\centering
	\caption{\textcolor{black}{EXISTING CONSUMER ELECTRONICS WORKS ON SMART HEALTHCARE}}
	\label{TABLE:existing_smart_healthcare}
	\begin{tabularx}{\textwidth}{@{} l C C C @{}}
		
		\hline \hline
		\addlinespace
		Existing Works & System Details &  Application to Smart Healthcare & Characteristics  \\
		\hline 
		\hline
		\addlinespace
		Ivanov, et al. 2012 \cite{Ivanov_IEEE-TCE_2012}& Cooperative wireless sensor network (WSN) with wireless body area network (WBAN)  & Healthcare monitoring includes ECG, blood oxygen level, and body temperature   & Energy efficient and cost effective   \\
		\addlinespace
		
		Spinsante, et al. 2012 \cite{Spinsante_IEEE-TCE_2012} & An integration of WSN framework, bluetooth, and digital TV  &Remote health monitoring and smart sensor & Simplistic and cost effective  \\
		\addlinespace
		Dey, et al. 2017 \cite{Dey_IEEE-TCE_2017} & Wireless sensor network (WSN) and Zigbee technology  &  ECG home healthcare monitoring & Improvement in device integration, reliability, and latency  \\
		\addlinespace
		Lee, et al. 2018 \cite{Lee_IEEE-TCE_2018} & Analog front end circuit and Digital signal processor & Arrhythmia monitoring & Enhancement in accuracy and low power consumption \\
		\addlinespace
		\textbf{Proposed System} & Seizure detection and control using IoMT & Remote EEG health monitoring and control of seizure & Accurate and  low power consumption \\
		
		\hline \hline
		
	\end{tabularx}
\end{table*}

Several seizure detection algorithms like wavelet decomposition \cite{Shoeb_IEEE-EMBS_2004},  phase coherence \cite{Mormann_Elsevier_2000}, and signal synchronization have been proposed. The implementations of these algorithms are only confined to powerful desktop computers and are not applicable to an implanted device. Over the last few years, considerable research has been  focused on developing implantable devices \cite{Raghu_JNE_2009, Bhavaraju_IEEE-TBE_2006, Salam_IEEE-TBCS_2011}. The event based algorithm \cite{Raghu_JNE_2009} relies on distributing EEG datasets into identical sized events. The threshold voltage associated with EEG abnormalities defines a seizure state.  The detector depends on both positive and negative threshold voltages, but can lead to false detection. The detection method based on support vectors \cite{Verma_IEEE-JSSC_2010} has improved the detection accuracy considerably but requires numerous support vectors to define a seizure and normal state. The drawback is high cost and high power consumption.  The detector in \cite{Patel_PCTHealth_2009} needs complex digital circuitry and an application specific chip to achieve the required sensitivity.  The preamplifier based detection technique in \cite{Salam_IEEE-TNSRE_2012, Salam_IEEE-TBCS_2011}, which is implemented with CMOS technology, is very useful for epileptic seizure detection but its power consumption is relatively high. There are also some noise related issues due to poor noise performance of CMOS technology \cite{Zhang_IEEE-TBCAS_2009}. In the current work, to mitigate the noise problem, we present a detection method which is energy efficient and less vulnerable to noise. Focal electrical stimulation \cite{Theodore_Neurology_2004, Boon_Neurotherapeutics_2009} has demonstrated promising results for some refractory patients where a closed loop responsive neuro-stimulator records EEG signals and triggers a stimulation upon seizure detection. Focal drug delivery \cite{Stein_Epilepsy_2000} enables drug injection directly onto the epileptogenic zone to disrupt seizure progression, improving the effectiveness of the medication. An asynchronous drug delivery system \cite{Salam_IEEE-TNSRE_2012} has been proposed for the treatment of refractory epilepsy which provides considerable reduction in power consumption.

\textcolor{black}{In our previous work \cite{Sayeed_ICCE_2018}, an SRA-based seizure detector was proposed which detects seizure with enhanced accuracy and provides low power consumption, but seizure control was not considered. In the current extended paper, an automated unified system is proposed in the IoMT platform, which performs both seizure detection and drug injection simultaneously to disrupt and control seizure propagation.}
\textcolor{black}{A different version of a drug delivery system \cite{Sayeed_ICCE_2019} has been proposed by the authors, which uses a machine learning classifier for seizure detection and an electromagnetic micro-pump for drug delivery. The system needs both software and hardware validation with extensive EEG data, and this is an ongoing project in our lab. However, the proposed cSeiz in the present work has been extensively validated, and uses a novel signal rejection algorithm (SRA) for seizure detection and a piezoelectric micro-pump for drug injection.}

\section{The Proposed Novel Drug Delivery System in the Internet of Medical Thing Perspective}
\label{Sec:System_Overview}

\begin{figure*}[htbp]
	\centering
	\includegraphics[width=0.997\textwidth]{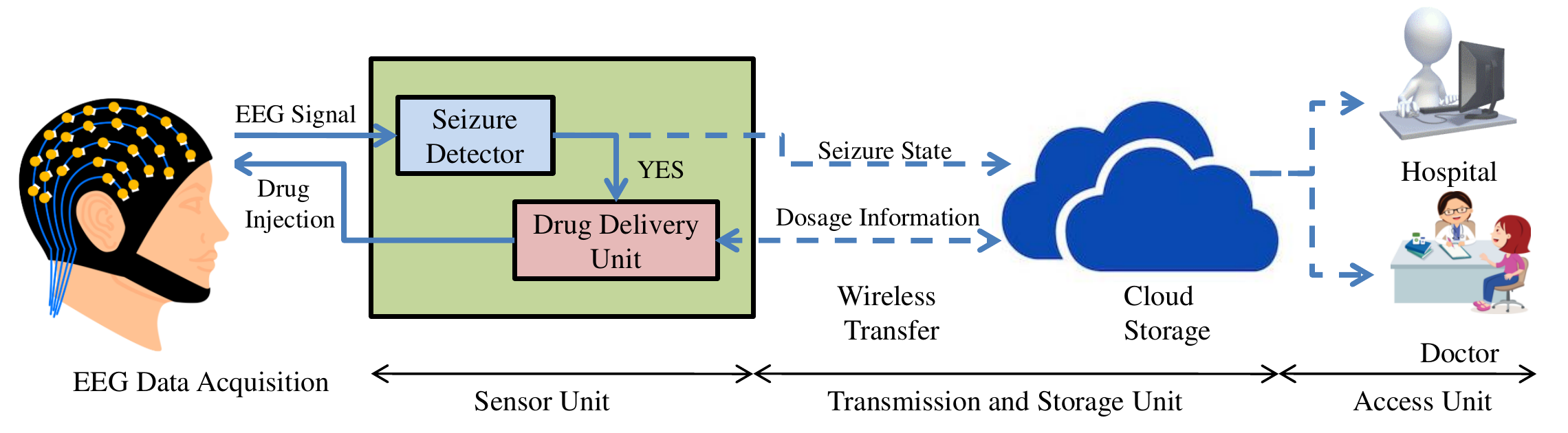}
	\caption{Proposed cSeiz in the Internet of Medical Things (IoMT).}
	\label{FIG:model_smart_DD_IOT}
\end{figure*}

Due to increased population, traditional healthcare systems are not able to provide necessary services to everyone. Smart healthcare utilizes the limited resource in an efficient way to fulfill everyone's needs \cite{Sayeed_ISES_2018}, \cite{Mohanty_IEEE-CEM_2016}. 
The IoMT in smart healthcare is an integration of universal communication and connectivity where all the necessary components can be connected together \cite{Ullah_ICISE_2016}, \cite{Sundaravadivel_IEEE-CEM_2018}. The proposed device is divided into three components, as shown in Fig. \ref{FIG:model_smart_DD_IOT}.

\subsection{Sensor unit}
The sensor unit consists of an EEG pre-processing unit, a seizure detector and the drug delivery unit. The input EEG signal is analyzed by a pre-processing unit. Upon seizure detection, drug injection takes place into the onset area. The information relating to patient's seizure state and dosage information is then sent to remote storage through a wireless transfer.

\subsection{Transmission and storage unit}
The transmission unit acts as an interface between the sensor unit and cloud storage. The main function of the storage unit is to store and manage the patient's data. Cloud storage is preferred as it enables data to be accessed from anywhere.

\subsection{Access unit}
This unit allows health professionals like physicians, health practitioners, and hospitals to access data from the cloud. The information relating to a patient is continuously stored in the cloud. In the case of a seizure, a notification is sent to the corresponding physician for proper action. The physician will check the patient's medication history as well as documented seizures and prescribe the required dosage for the treatment of epilepsy. The data is also accessible to the patients which allows them to be updated with current health conditions \cite{Ullah_ICISE_2016, Sayeed_ISC2_2018, Sayeed_TCE_2019-Aug_Neuro-Detect}.

\begin{figure*}[htbp]
	\centering
\subfigure[Flowchart]{\includegraphics[width=0.50\textwidth]{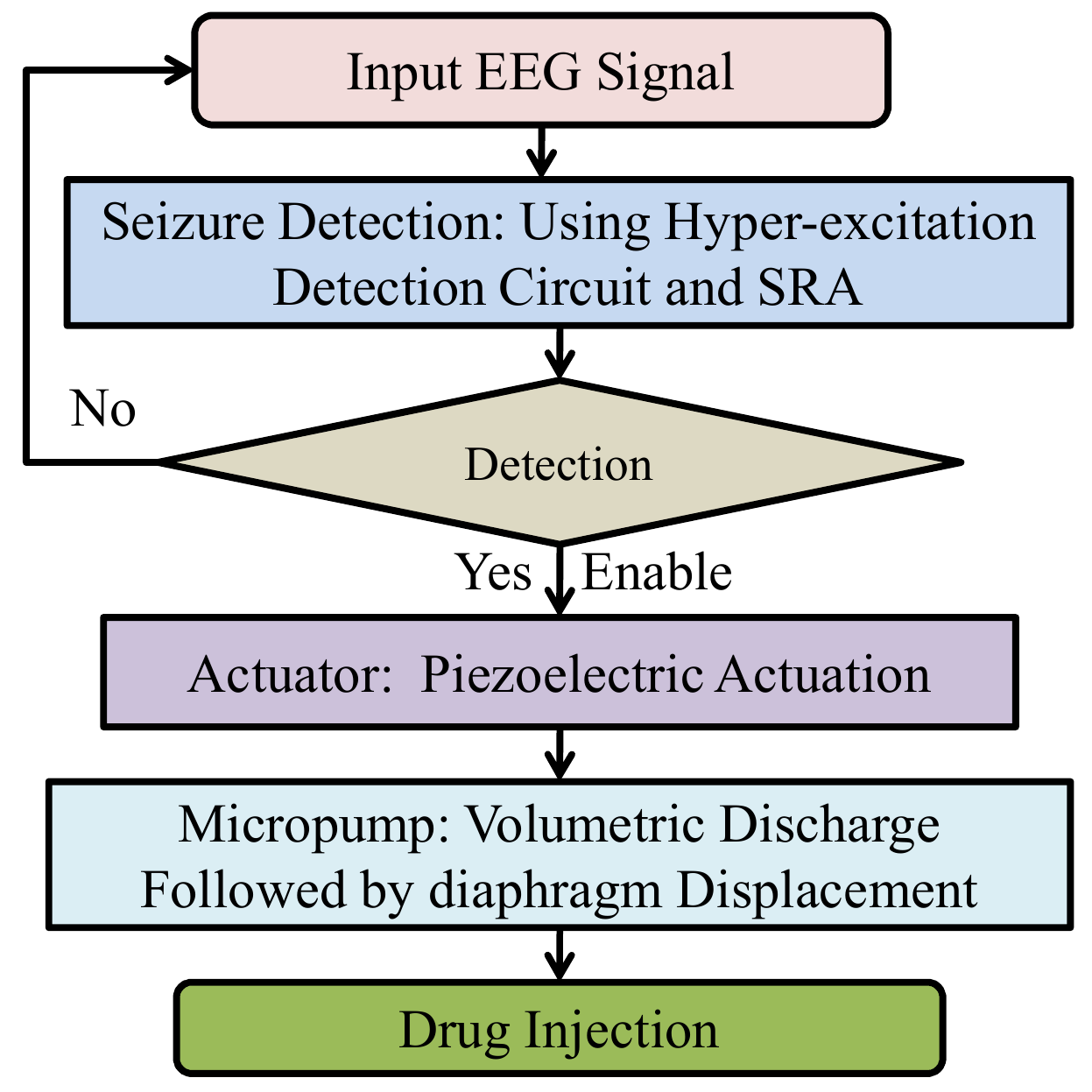}\label{fig:flowchart_drug_delivery_system}}
\subfigure[Architecture]{\includegraphics[width=0.89\textwidth]{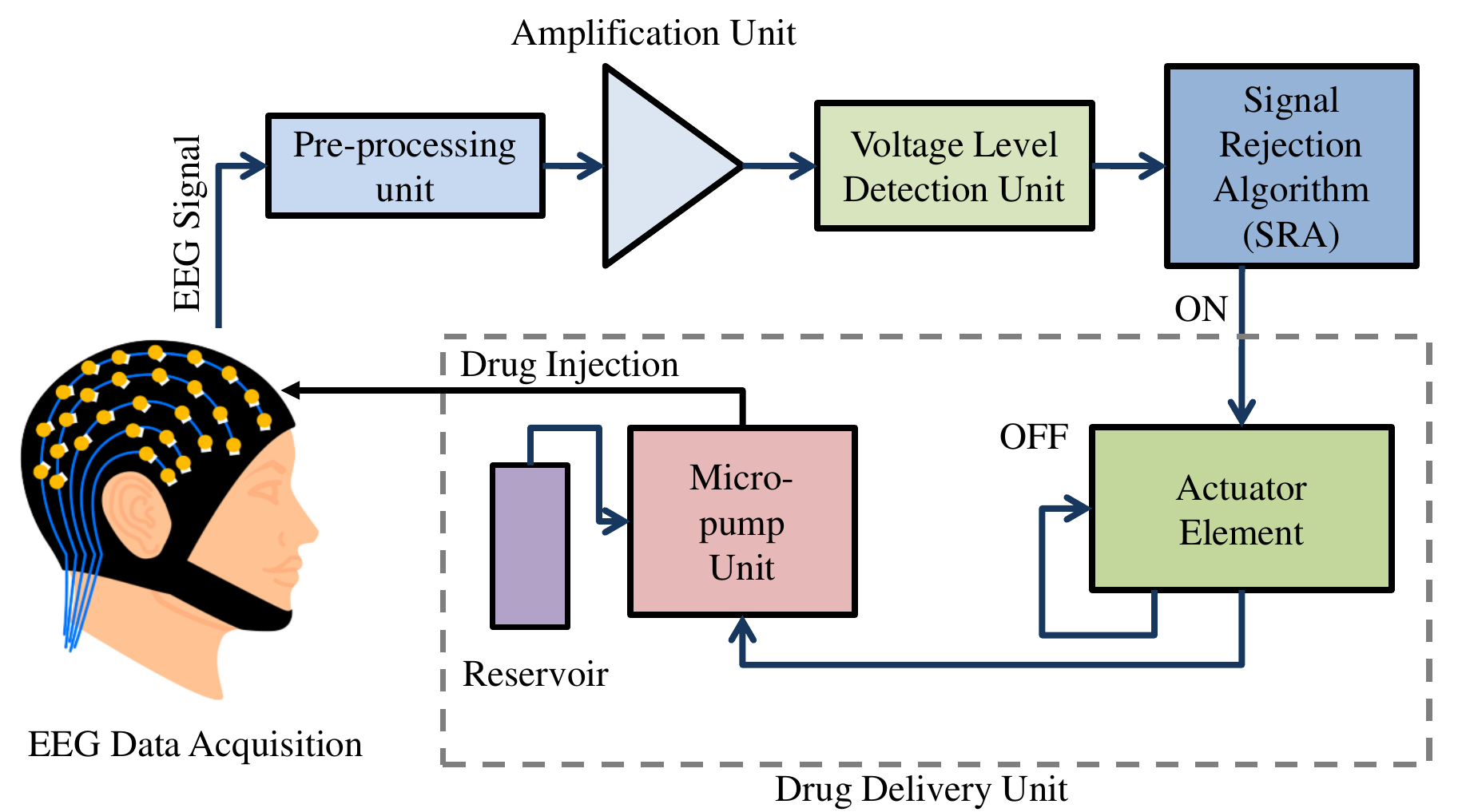}\label{fig:architecture_proposed_system}}
%
			\caption{Proposed Drug Delivery System (a) Flowchart (b) Architecture.}
			\label{fig:drug_delivery system}
\end{figure*}

The EEG signals are initially pre-processed through filters which are then passed through an amplifier and voltage level detector. The seizure detector analyzes and removes unwanted signals from the output of the voltage level detector.	Upon seizure detection, the actuator enables the micro-pump to inject a dose into the onset area, which leads to a disruption of seizure progression. The reservoir supplies the required drug to the micro-pump unit. Fig. \ref{fig:drug_delivery system} illustrates the flowchart and architectural diagram of the proposed system.

\section{The Proposed Novel Seizure Detector}
\label{Sec:Proposed_Detector}



\begin{figure}[htbp]
	\centering
	\includegraphics[width=0.65\textwidth]{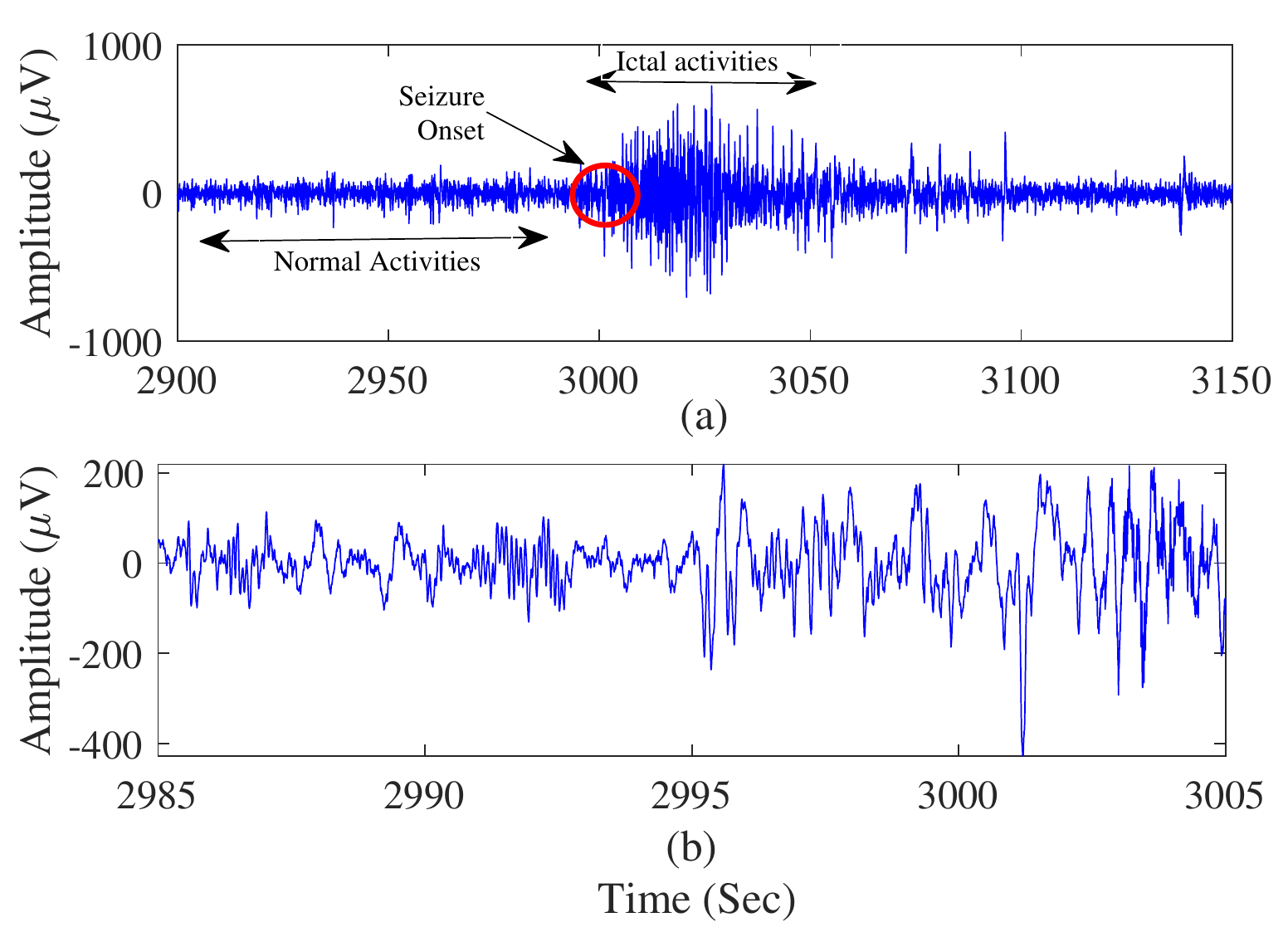}
	\caption{Seizure Activity Characterization in the Time Domain (a) Invasive Electroencephalography (EEG) of an Epileptic Seizure (b) Zoom Inset 2985-3005 seconds.}
	\label{FIG:eeg_signal}
\end{figure}

\begin{figure}[htbp]
	\centering
	\includegraphics[width=0.99\textwidth]{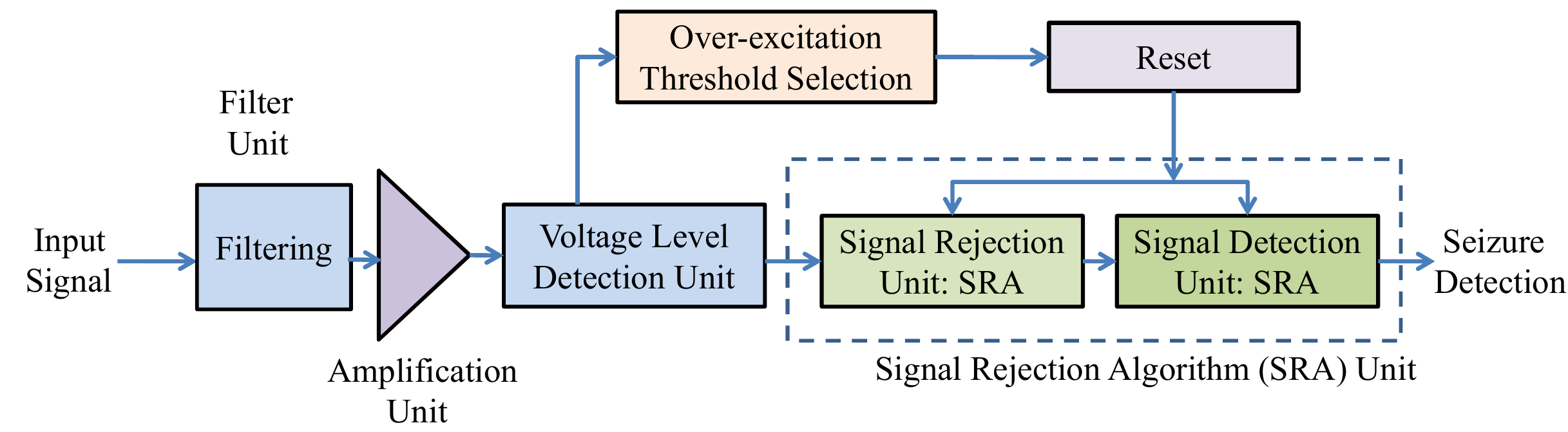}
	\caption{Architecture of the Proposed Seizure Detector.}
	\label{FIG:architecture_seizure_detector}
\end{figure}

\begin{figure}[htbp]
	\centering
	\includegraphics[width=0.56\textwidth]{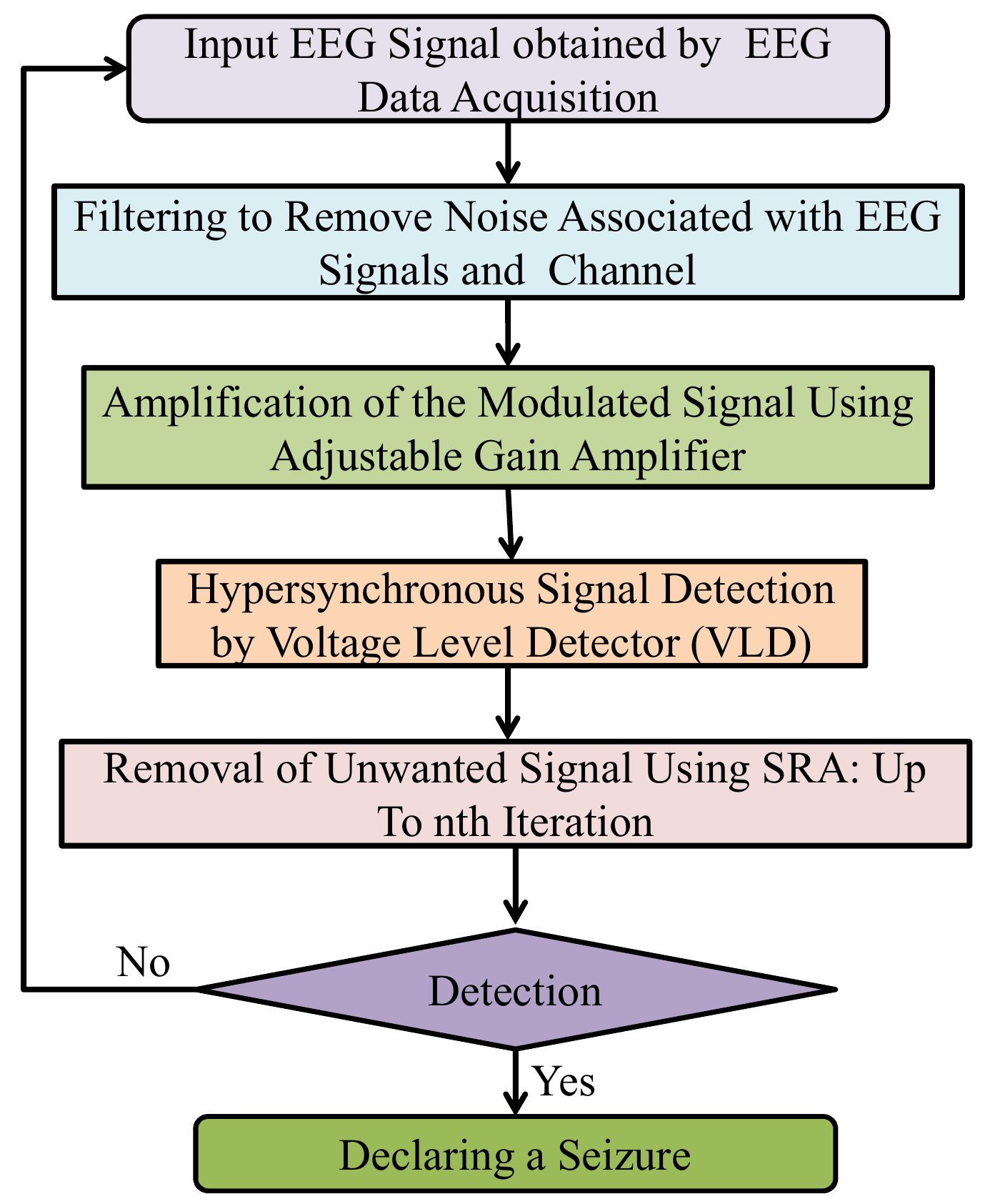}
	\caption{The Proposed Steps for the Seizure Detection in cSeiz.}
	\label{FIG:flowchart_seizure_detector}
\end{figure}

\textcolor{black}{The proposed seizure detector (SD) monitors the brain activity at the seizure onset area. Fig. \ref{FIG:eeg_signal} shows a characterization of seizure onset. The architecture and flowchart of the proposed detector are shown in Fig. \ref{FIG:architecture_seizure_detector} and Fig. \ref{FIG:flowchart_seizure_detector}, respectively. The input EEG signals are filtered and submitted to an amplification unit. The amplified signals of desired range are then passed through a voltage level detector (VLD). The resulting hyper-synchronous pulses from the VLD are then submitted to the signal rejection algorithm (SRA) unit. The SRA unit eliminates unwanted signals and noise. The elimination of unnecessary signals continues until the number of hyper-synchronous pulses surpasses the threshold value. Seizure detection is characterized by the following expression:}
  \begin{equation}
  V_{SE}(n)=\begin{cases}
  1, & \text  {seizure,  for $V(n-i)=1 \cdots\; \text{and}\; V(n)=1$}\\
  0, & \text{no seizure, otherwise,}
  \end{cases}
  \label{eq:seiz}
  \end{equation}
where, $i = 1, 2, 3, \dots, N$. 

\subsection{Hyper-synchronous Signal Detection Circuit}

\begin{figure}[!h]
	\centering
	\includegraphics[width=0.75\textwidth]{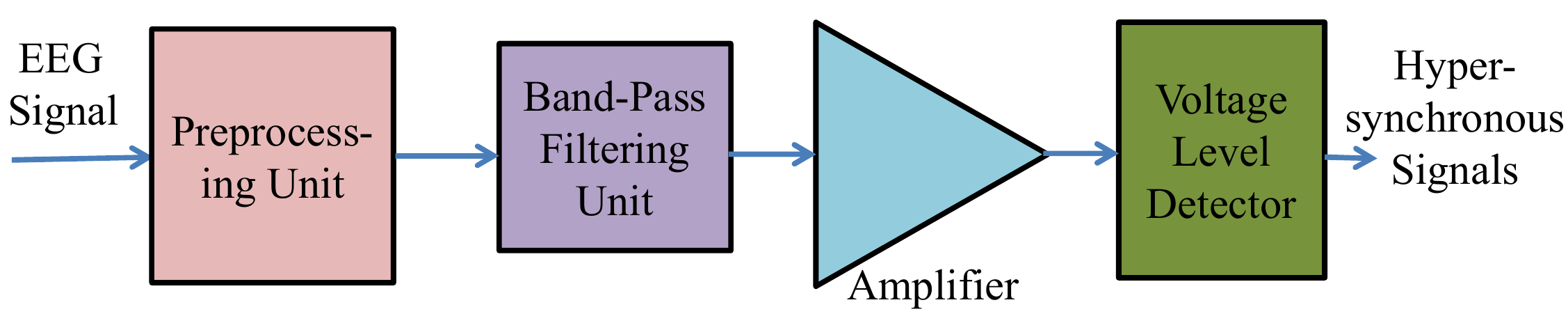}
	\caption{\textcolor{black}{Hyper-synchronous Signal Detection Circuit}}
	\label{FIG:hypersynchronous_circuit}
\end{figure}

\textcolor{black}{The proposed circuit in Fig. \ref{FIG:hypersynchronous_circuit} consists of a band pass filter, an amplifier, and a voltage level detector (VLD). Band pass filter eliminates the unwanted signals and noise associated with the scalp-EEG signals and only keeps signals of desired frequency range. The low amplitude neural signals need to be amplified prior to analysis \cite{Salam_BioCAS_2010}. The desired level of the signals is achieved by the amplification unit. The VLD analyzes the amplified signals and detects hyper-synchronous pulses. The threshold values ($V_{max}$, $V_{min}$) of the VLD are determined from the analysis of known seizure instances. The detection of the hyper-synchronous signal is performed based on the following  equation \cite{Salam_BioCAS_2010}:}
\begin{equation}
V_{vld}(n)=\begin{cases}
1, & \text  {  for $V_{max}>V_{mod}(n)>V_{min}$}\\
0, & \text{ otherwise.}
\end{cases}
\label{eq:detect}
\end{equation}

\subsection{Signal Rejection Algorithm (SRA): Detection of seizure onset from hypersynchronous signals}

\textcolor{black}{The hyper-synchronous signals from the VLD  are analyzed  and spurious pulses are eliminated using SRA. The elimination of unwanted signals is performed by:}
\begin{equation}
V_{SE1}(n)=\begin{cases}
0, &  V(n-2)=0\; \text{or}\;V(n-1)=0\\
V(n), & \text{otherwise.}
\end{cases}
\label{eq:3}
\end{equation}
The spurious pulse is further eliminated using:
\begin{equation}
V_{SE2}(n)=\begin{cases}
0, & V(n-1)=0 \; \text{and}\;V(n)=1\\
V(n), & \text{otherwise.}
\end{cases}
\label{eq:4}
\end{equation}
The seizure onset is characterized by:
\begin{equation}
V_{SE}(n)=\begin{cases}
\text{Seizure}, & V(n-2)=1 \; \text{and} \; V(n-1)=1,\\
&  \hspace{0cm} \text{if} \; V(n)= \text{1 or 0}  \\
0, & \text{otherwise.}
\end{cases}
\label{eq:6}
\end{equation}


\textcolor{black}{Neural signals are continuously monitored and seizure is detected from the hyper-synchronous pulses ($V{vld}$). Within a time frame, the unwanted pulses are eliminated if their amplitude is lower than the threshold value. A seizure onset is declared when the SRA completes $n$-th iteration. If the number of hyper-synchronous pulses is greater than the threshold number, a seizure is declared according to equations (\ref{eq:3}), (\ref{eq:4}), and (\ref{eq:6}).}

\section{Proposed Drug delivery subsystem}
\label{Sec:Proposed_Delivery-System}

Considerable research is focused on developing microfluidic  systems for drug delivery. Micro-pumps are an integral part of a microfluidic system, which controls small fluid volumes. Several micro-pumps have been proposed based on different actuation mechanisms including electrostatic, electromagnetic, electro-osmetic and piezoelectric. Piezoelectric actuation leads to a higher deflection, faster response, and higher actuation force as compared to other mechanisms  \cite{Singh_Sensors-actuators_2015}. The elimination of the valves makes the pump more reliable and simpler \cite{He_JSV_2017, Kawun_Sensors-actuators_2016, Sayeed_ICCE_2019}. The working principle of the micro-pump is based on higher pressure loss in the nozzle direction rather than in the diffuser direction. The micro-pump works in two different modes: the supply mode and the pump mode. During the supply mode, the actuator deflects the diaphragm which increases the volume of the pump chamber. The lower pressure loss across the inlet element, which acts as a diffuser, enables a larger volume to be transferred  to the pumping chamber using an inlet port than through the output. During the pump mode, the actuator decreases the volume of the pumping chamber. The lower pressure loss across the outlet element, which acts as a diffuser, enables a larger volume to be transferred to the pumping chamber using outlet port than through the input. As a result, a net volume is transported from the inlet side to the outlet side.
Once a seizure occurs, the drug delivery unit is turned on. A voltage is applied to the piezoelectric disc for actuation, which deflects the diaphragm for the pumping action. The supply mode leads to the increment of the pump chamber volume whereas the pump mode reduces the chamber volume.  The reservoir provides the required drug to the pump chamber when necessary. Finally, the required dosage is obtained from the nozzle geometry and operating conditions and is then injected to the epileptogenic zone for treatment. 

\subsection{Design of the valveless micro-pump}

\begin{figure}[htbp]
	\centering
	\includegraphics[width=0.75\textwidth]{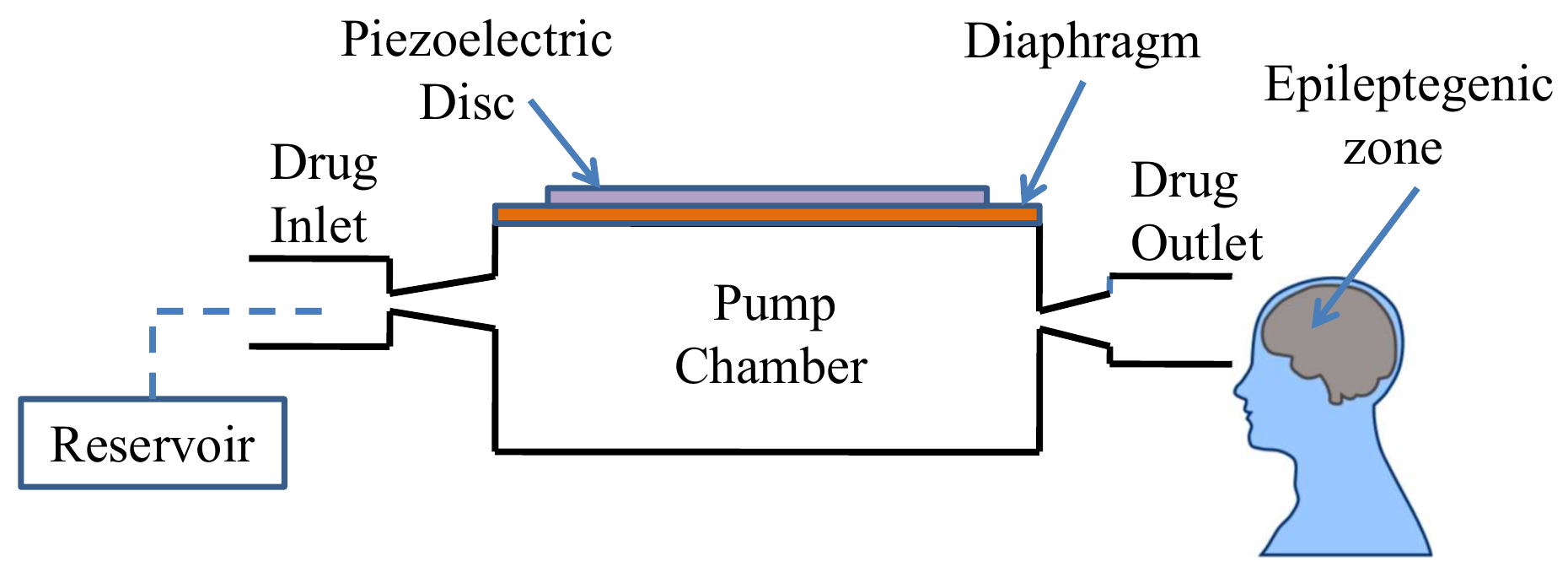}
	\caption{A Valveless Micro-pump Explored for Use in cSeiz.}
	\label{FIG:design_valveless_micropump}
\end{figure}

\begin{figure}[htbp]
	\centering
	\includegraphics[width=0.48\textwidth]{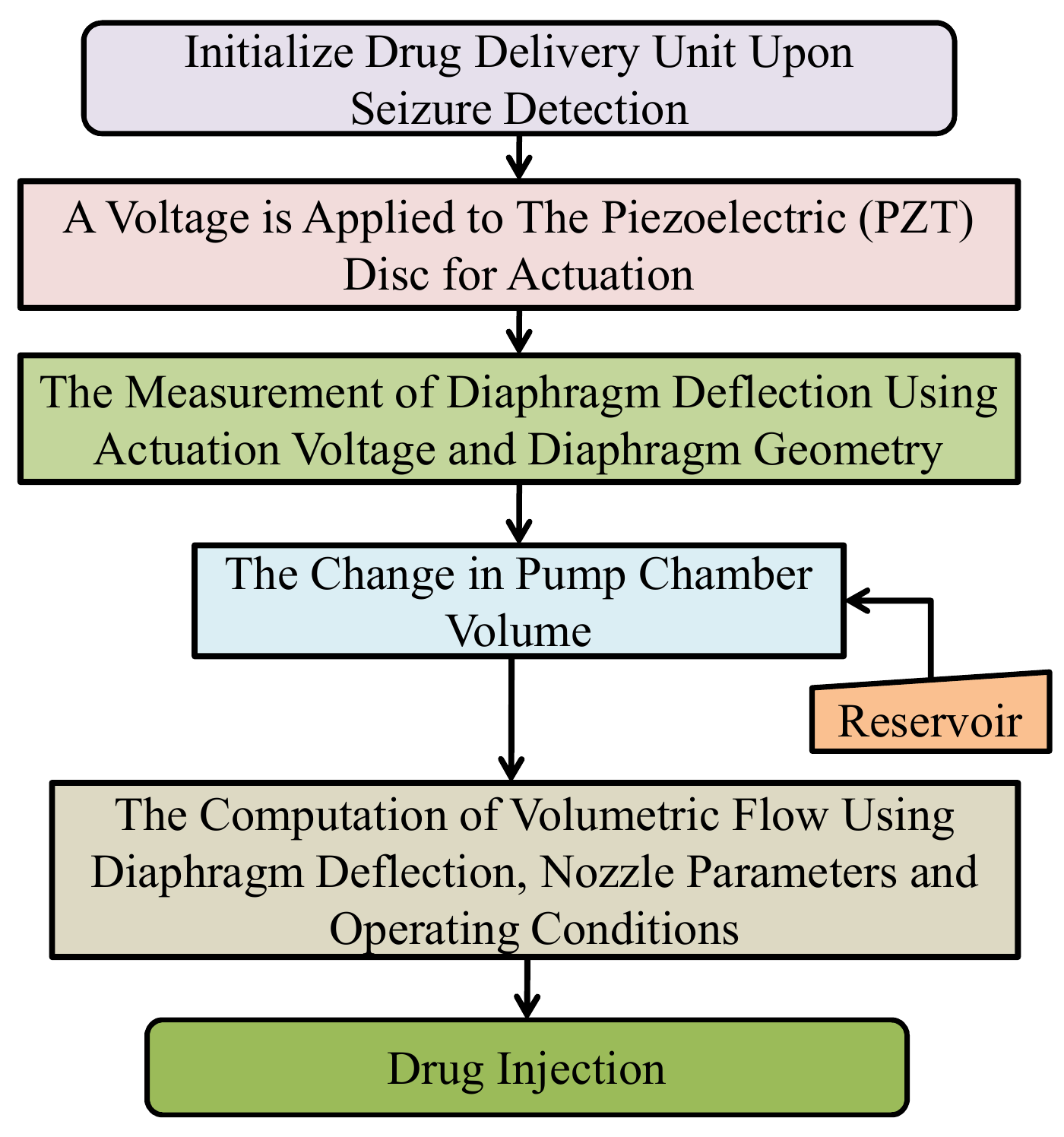}
	\caption{The Steps for Drug Injection in cSeiz.}
	\label{FIG:flowchart_drug_delivery_unit}
\end{figure}

A schematic diagram of the valveless micro-pump \cite{Kawun_Sensors-actuators_2016}, \cite{Ramaswamy_ISDMISC_2011} is shown in Fig. \ref{FIG:design_valveless_micropump}, while Fig. \ref{FIG:flowchart_drug_delivery_unit} shows the steps followed during drug injection. 

The diaphragm is operated by the actuator element. An electric field on a PZT disc leads to a displacement in the vertical direction, which is necessary for pumping action.  The diaphragm or pump membrane, which is attached to the PZT disc, covers the hole of the pump chamber. In the pumping mode, the flow rate is a function of stroke volume which depends on the diaphragm motion. The applied excitation voltage to the actuator oscillates the diaphragm and provides the required displacement \cite{Singh_Sensors-actuators_2015, Kawun_Sensors-actuators_2016}. Stainless steel is a useful material for building pump chambers as it has excellent biocompatibility and good corrosion resistance. The pump chamber has been designed in such a way that it corresponds to the design requirement for the liquid pump along with the actuator dimension. The displacement of the diaphragm is proportional to the applied driving voltage. A DC to AC conversion and amplifications are performed to obtain the driving voltage. The DC supply voltage is 5 V and the driving output voltage is 20 V.







\subsection{Model of the micro-pump}

The diaphragm displacement is calculated from:  
\begin{equation}
X= \frac{0.55 R^2 F}{E {t_d}^3},   
\end{equation}
where $R$, $t_d$, and $E$ are the radius, thickness, and Young Modulus of the diaphragm, respectively. The force produced by piezoelectric actuation is denoted by $F$, and is inversely proportional to the stress constant of the corresponding material and directly proportional to the applied voltage \cite{Ramaswamy_ISDMISC_2011}. The pressure loss coefficients across the diffuser and the nozzle are denoted by ${\epsilon}_{dif}$ and ${\epsilon}_{noz}$ \cite{Stemme_Sensors-actuators_1993}.
The volume flow per stroke can be calculated using the following:
\textcolor{black}{
\begin{equation}
V_{str} =\frac{2\pi}{3}X{R^2}.
\end{equation}}
It is assumed that the pressure loss coefficients in the diffuser and the nozzle are constant. The outlet flow has been measured in both supply and pump mode. The net volume flow during one complete pump stroke is obtained by \cite{Olsson_Sensors-actuators_1993}:
\textcolor{black}{
 \begin{equation}
Q=2  V_{str} f   \frac {(\sqrt{\eta}-1)}{(\sqrt{\eta}+1) },
\end{equation}}
\noindent where $f$ is the pump frequency and  $\eta=\epsilon_{noz}/\epsilon_{dif}$. $\eta$ is an important parameter which should be kept as large as possible to have a maximum volume flow.

\section{Consumer Electronics (CE) Proof of concept of the proposed cSeiz }
\label{Sec:Proposed_Modeling}

 The EEG signal is initially pre-processed through a band pass filter. The low amplitude neural signal is then amplified using an adjustable gain amplifier. The maximum and minimum voltages associated with the voltage level detector define the hyper-synchronous signal. The proposed detector extracts the seizure onset using SRA. Upon seizure detection, the piezoelectric disc deflects the diaphragm. The micro-pump injects drug onto the epileptegenic zone and disrupts seizure propagation.

\begin{figure*}[htbp]
\centering
\subfigure[Proposed drug delivery system]{\includegraphics[width=0.98\textwidth]{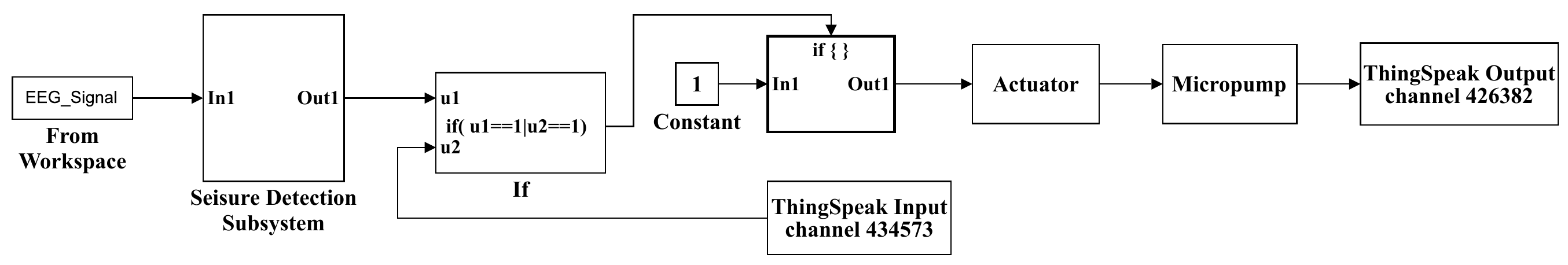}\label{fig:simulink_drug_delivery_system}}
\subfigure[Seizure Detection Subsystem]{\includegraphics[width=0.98\textwidth]{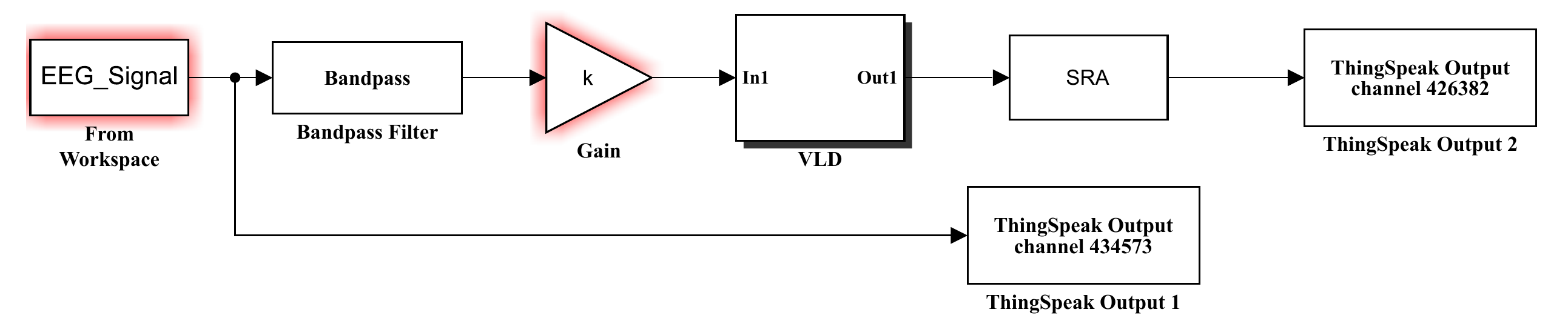}\label{fig:simulink_seizure_detector}}
\subfigure[Pattern-Independent Power Measurement Setup]{\includegraphics[width=0.98\textwidth]{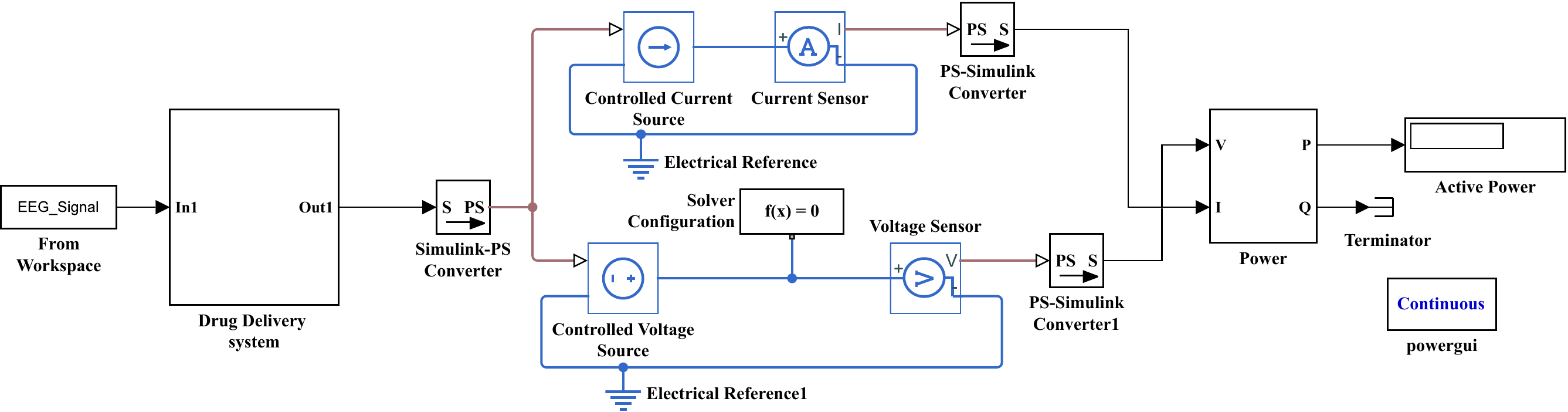}\label{fig:simulink_power_measurement}}
\caption{System-Level Simulator Model of (a) Proposed drug delivery system.  (b) Seizure Detection Subsystem. (c) Pattern-Independent Power Measurement Setup.} 
\label{FIG:drug_delivery_system}
\end{figure*}



	

System-level prototyping of the proposed cSeiz is performed as a first step towards CE prototyping. The system level model of the proposed system is shown in Fig. \ref{FIG:drug_delivery_system}. EEG signals are fed into the system.  A band pass filter eliminates unwanted noise and extracts all seizure onset information. The adjustable gain amplifier enables signals to be amplified to the desired level.  Hyper-synchronous signals are detected by the VLD. If the voltage is within the range, the function outputs a 1, otherwise it is zero. 
\textcolor{black}{The maximum and minimum voltage of the VLD is determined by heuristic analysis of the amplified signal. For an epileptic subject of n number of seizures; where n denotes the even number, the $V_{max}$ and $V_{min}$ values are computed from $n/2$ seizure instances. If $n$ is odd, the $V_{max}$ and $V_{min}$ values are obtained from $(n-1)/2$ seizure instances. The average optimal values have been adjusted by trial and error method, which are then applied to unknown seizure and non-seizure instances. The hyper-synchronous signal which is obtained from VLD is given to SRA unit. In the first iteration of signal rejection, if a sample is either '0' or '1' and the previous two sample is '0', the algorithm outputs a 0. In the next iteration, if the previous sample is '0', the algorithm outputs a '0'. For seizure onset formation at the $(n-k)$th iteration, if the previous sample is '1', the algorithm results a '1'. In the $n$th stage, the algorithm define a seizure onset if the number of hyper-synchronous pulses which is denoted by '1' exceeds the threshold value. The optimal value of threshold, $n$, and $k$ can be achieved by heuristic analysis of the known seizure and non-seizure instances. The SRA technique has been conceptualized in Table \ref{TABLE:SRA_technique}, where the signals have been analyzed for seizure onset and non-seizure onset instances.}

\begin{table}[htbp]
	\centering
	\caption{SIGNAL REJECTION AlGORITHM (SRA) TECHINIQUE.}
	\label{TABLE:SRA_technique}
\begin{tabular}{lc@{\hspace{1.1\tabcolsep}}c}
			 \hline \hline
			 \addlinespace
		 Location & Normal EEG & Seizure Onset\\
		 
		\hline \hline
		\addlinespace	
		VLD output 
		&0 1 0 0 1 1 0 0 0  & 1 1 1 0 1 1 1 \\
		SRA ($1^{st}$ iteration)
		&0 1 0 0 0 1 0 0 0  &0 1 1 0 1 1 1 \\
		SRA ($2^{nd}$ iteration)
		&0 1 0 0 0 0 0 0 0  &0 0 1 0 0 1 1 \\
		SRA ($3^{rd}$ iteration)
		&0 0 0 0 0 0 0  0 0&  0 0 1 1 0 1 1 \\
		SRA ($n^{th}$ iteration)
		&0 0 0 0 0 0 0 0 0  &0 0 1 1 1 1 1 \\
		Seizure Detection
		& 0 & 1 \\
		\hline  \hline
	\end{tabular}
\end{table}

The mathematical model of the micro-pump unit was implemented using user defined functions which are available in the system level simulator. Diaphragm deflection has been calculated using different parameters such as voltage, actuation frequency, diaphragm geometry, and material properties which are input to the actuator. However, volumetric discharge has been computed from the diaphragm deflection, nozzle parameters and other operating conditions. An input pattern independent method is employed for the accurate estimation of power dissipation \cite{Albalawi_ISVLSI_2016}. In this method, the proposed design is simulated with different EEG datasets of identical size. The average of the simulation results is considered as an estimate for power dissipation. The sensors and measurement blocks of the system level simulator extract the voltage and current values from the design, which is viewed as a black box,  and calculates power dissipation.

\begin{figure}[htbp]
	\centering
	\includegraphics[width=0.55\textwidth]{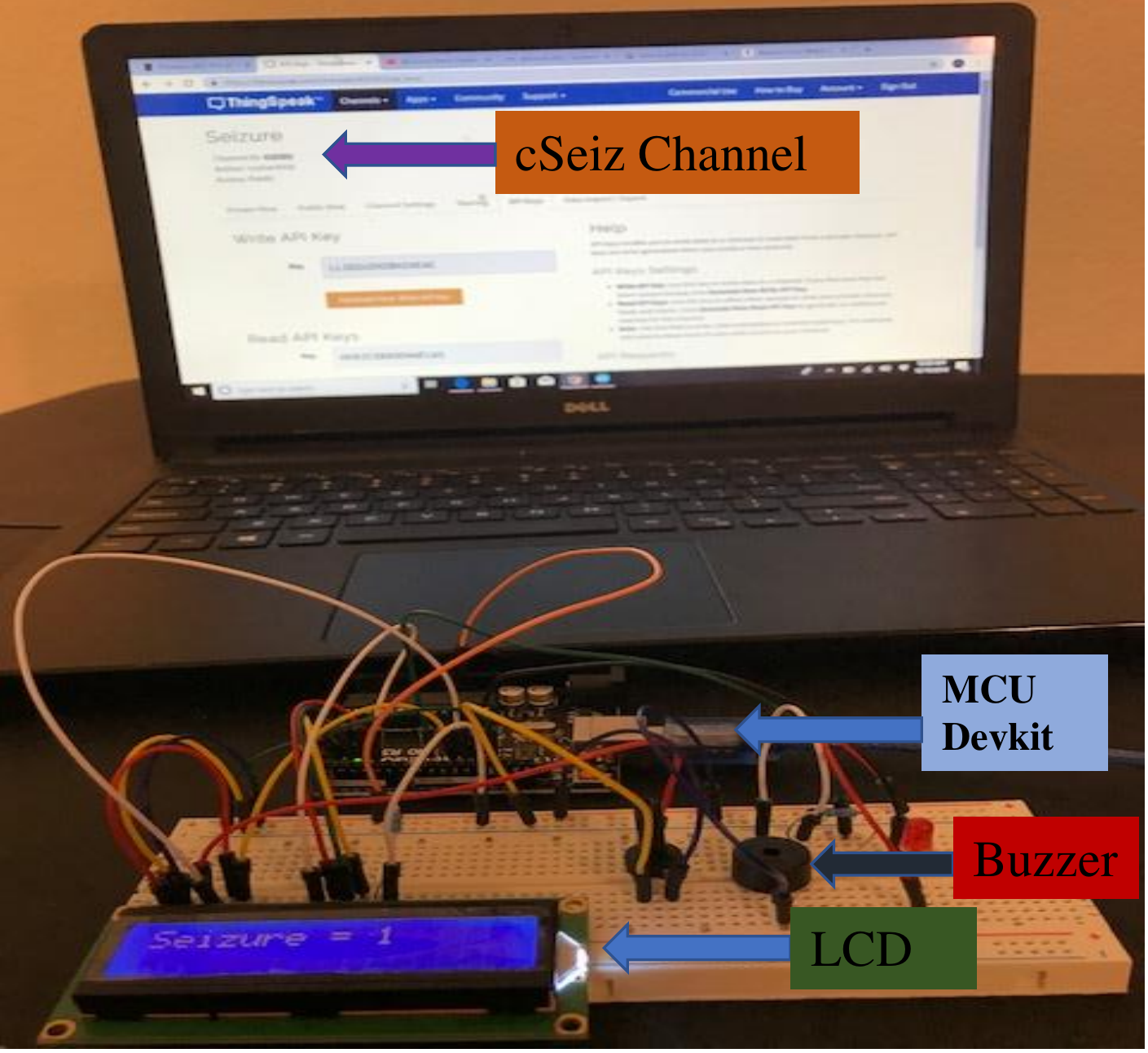}
	\caption{CE Prototyping of the Proposed cSeiz Device.}
	\label{FIG:hardware_prototype}
\end{figure}

A hardware-in-the-loop simulation approach was followed for the CE prototyping of the proposed system. A vendor provided hardware support package was used in the system level simulator and the proposed model was run on the actual board.  EEG data, seizure state and dosage information are continuously stored on the cSeiz channel in the open  IoT cloud.  A Liquid Crystal Display (LCD), which is attached to the board, shows information about seizure state and drug dosage. If any seizure occurs, a notification is sent to the designated user through the cloud. The EEG data and dosage information are sent to open cloud storage, while the system concurrently receives dosage information prescribed by the physician. Both patients and medical professionals have access to the IoT cloud as well as the database using a REST API \cite{Sayeed_ISES_2018}. Fig. \ref{FIG:hardware_prototype} shows the CE prototyping of the proposed cSeiz. 

\section{Experimental Results}
\label{Sec:Experimental_Results}

The continuous and long term EEG recordings is taken from CHB-MIT scalp EEG database \cite{Golberger_physionet_2000}, \cite{Shoeb_thesis_2009}, which consists of EEG recordings from pediatric subjects. This work uses common EEG data from the following subjects: chb01, chb03, chb05, chob08, chb11, chb17, and chb19. EEG recording were sampled at 256 samples per second with 16 bit resolution. EEG electrode was placed based on International federation of Clinical Neurophysiology 10-20 placement system. Initially EEG signal is passed through a band pass filter of frequency range between 3 Hz to 29 Hz, which is  amplified to a certain level. The amplified signal is applied to VLD. The seizure onset information is extracted  using the VLD. The maximum ($V_{max}$) and minimum ($V_{min}$) voltages are computed by heuristic analysis of the known instances. As a good number of non-seizure signals fall on the VLD category, VLD produces a number of unwanted pulses. The output of the VLD, namely hyper-synchronous pulses ($V_{vld}$), are fed into the SRA. The SRA eliminates unwanted pulses in every iteration. The SRA iterations continue until the number of hyper-synchronous pulses surpass the threshold  number. The unwanted pulses are being eliminated as SRA completes the $n$-th iteration. The time frame ($T_f$) is in the range of milliseconds to seconds. For the patient specific detection, the value of $T_f$ , $V_{max}$, and $V_{min}$ can be varied according to the patients. It is also reported that some unwanted signals with amplitude of the VLD range make false detection. In order to solve this problem, statistical energy \cite{Sayeed_ISES_2018} in each time frame is calculated for the known seizure and non-seizure instances and optimal value is determined by heuristic approach as discussed earlier. The optimal threshold value of energy in each time frame is considered with the SRA unit. Even if a instance is incorrectly detected by SRA unit, the further analysis with threshold energy provides correct detection. The average amplitude of the seizure pattern at onset is between 150 mV and 450 mV.  The frequency range for epileptic discharge is between 3 and 29 Hz. The performance of the detector is measured using sensitivity, specificity, and latency.
	The sensitivity and specificity of the classifier were evaluated as follows:
	\begin{equation}
	\text{Sensitivity}=\frac{\text{True Positive}}{\text{True Positive} + \text{False Negative}}
	\end{equation}
	\begin{equation}
	\text{Specificity}=\frac{\text{True Negative}}{\text{True Negative} + \text{False Positive}}
	\end{equation}
	Seizure detection delay is called latency, which is the delay between the expert marked seizure onset and the seizure onset marked by the seizure detector.

\begin{figure}[htbp]
	\centering
	\includegraphics[width=0.6\textwidth]{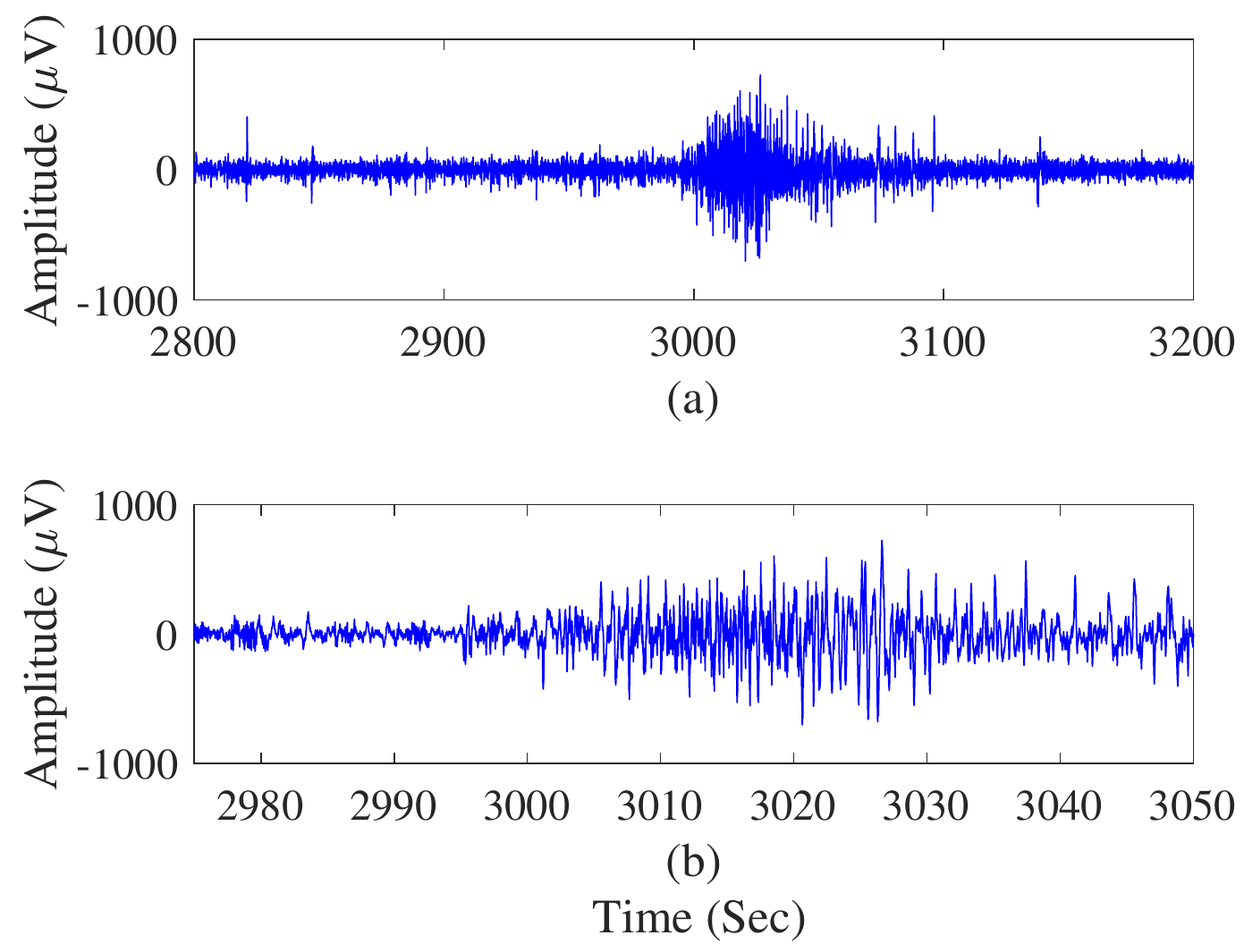}
	\caption{\textcolor{black}{ Transient Analysis. (a) Input EEG signal of 2800-3200 seconds (b) EEG signal of 2975-3050 seconds    }}
	\label{FIG:original_signal}
\end{figure}

\begin{figure}[htbp]
	\centering
	\includegraphics[width=0.6\textwidth]{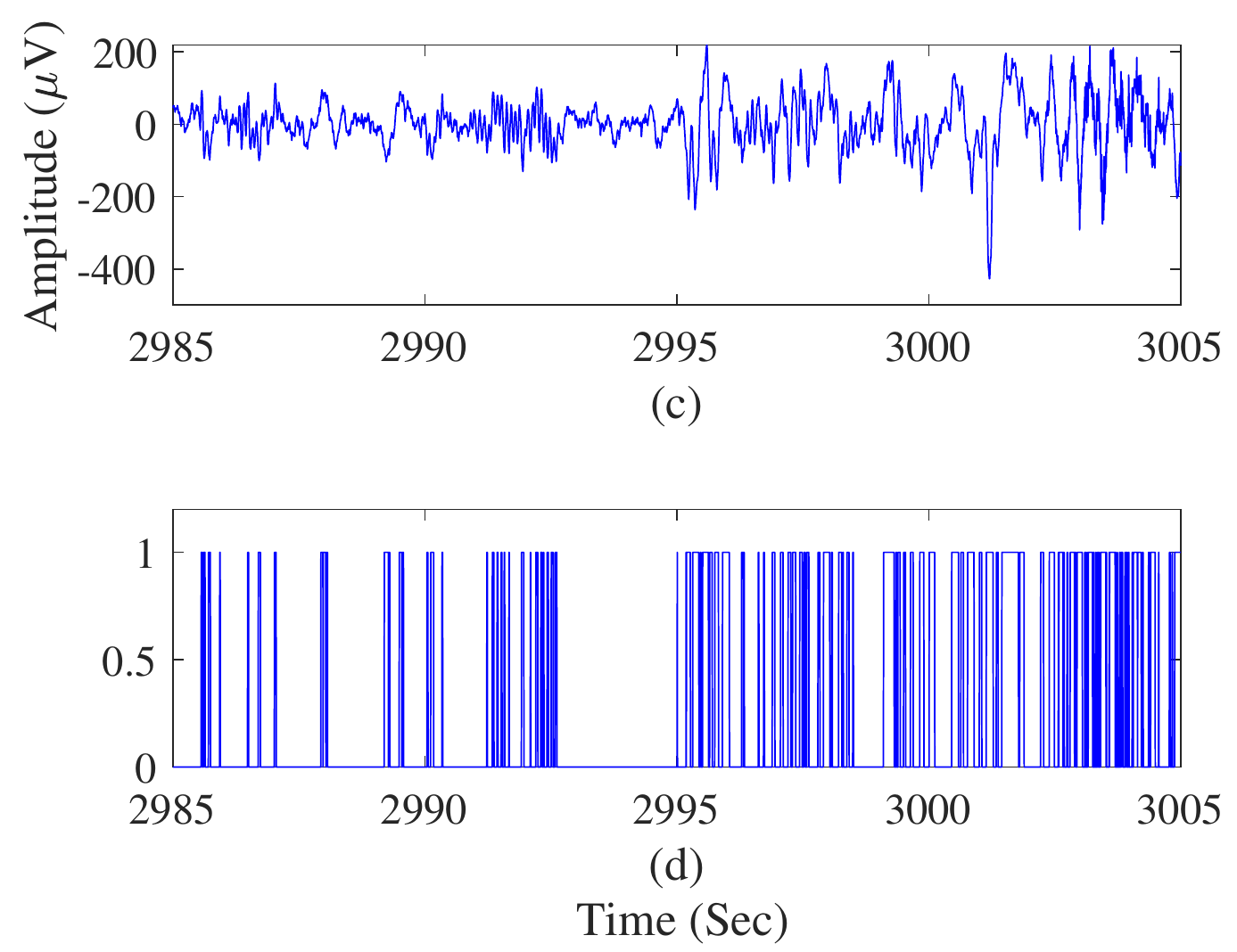}
	\caption{\textcolor{black}{ Transient Analysis (c) Zoom 2985-3005 seconds of input signal (d) Output of VLD at 2985-3005 seconds }}
	\label{FIG:VLD_waveform}
\end{figure}

\begin{figure}[htbp]
	\centering
	\includegraphics[width=0.6\textwidth]{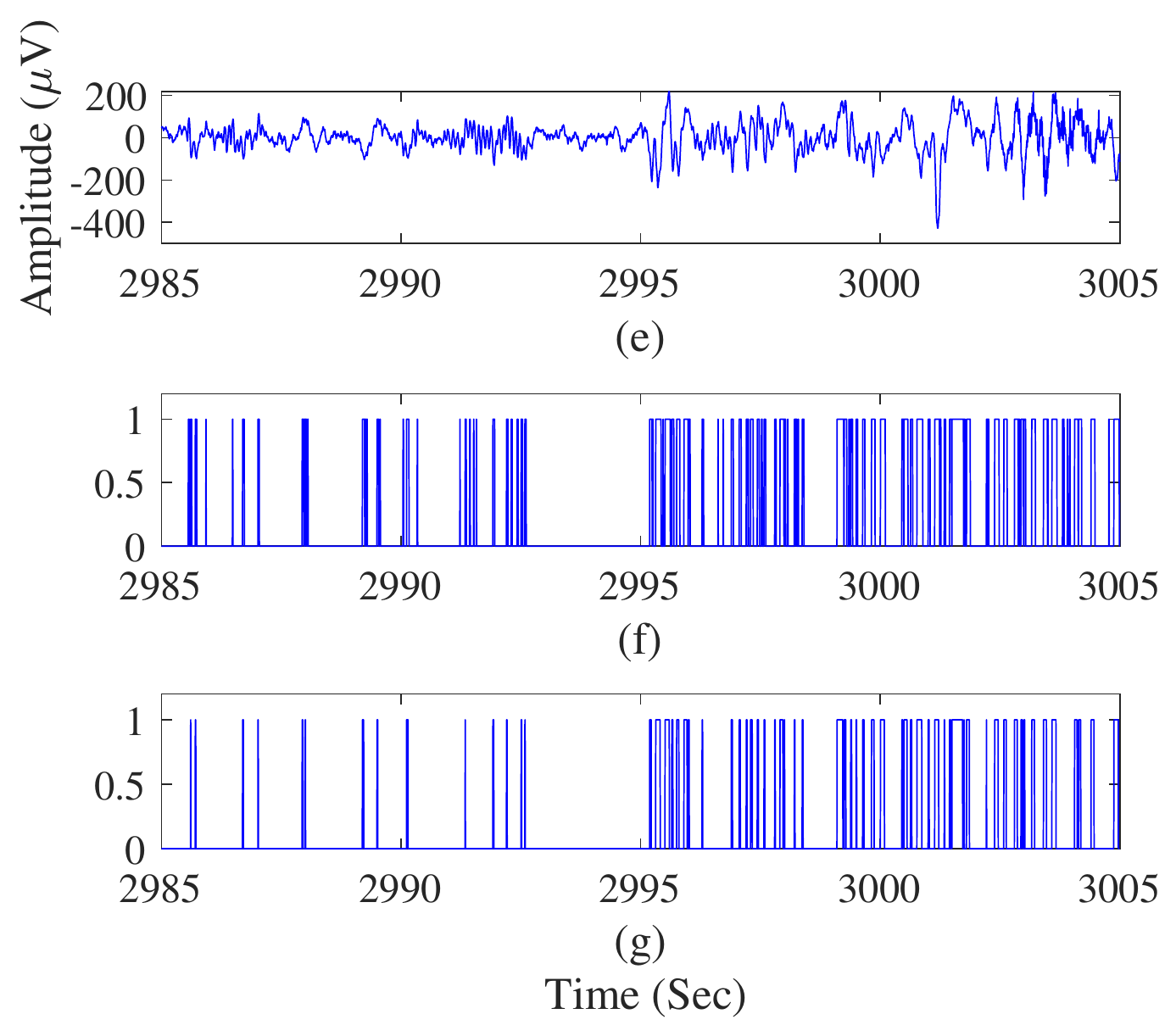}
	\caption{\textcolor{black}{ Transient Analysis (e) Zoom 2985-3005 seconds of input signal (f) Output of SRA after 1st iteration (g) SRA output after 2 nd iteration }}
	\label{FIG:SRA_12_waveform}
\end{figure}

\begin{figure}[htbp]
	\centering
	\includegraphics[width=0.6\textwidth]{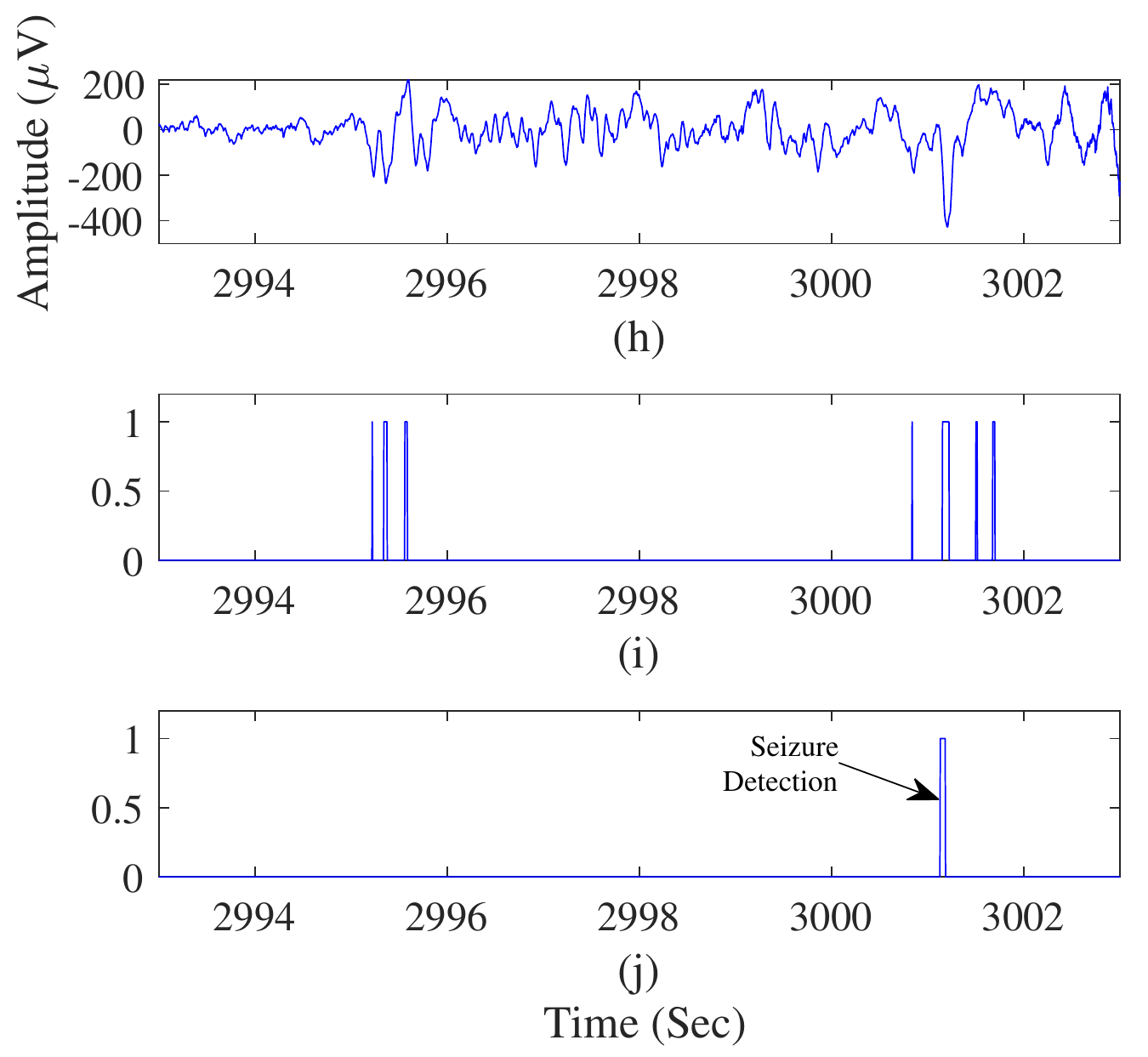}
	\caption{\textcolor{black}{ Transient Analysis (h) Zoom 2993-3003 seconds of input signal (i) SRA output after $(n-k)$th iteration (j) SRA output after $n$th iteration. }}
	\label{FIG:SRA_final_waveform}
\end{figure}

Figure \ref{FIG:original_signal}-\ref{FIG:SRA_final_waveform} shows the analysis of the EEG signal and detection of seizure for epileptic subject 1 (chb01), in which seizure starts at 2996 s and ends at 3036 s. The time frame for chb01 is selected as 500 ms. Figure \ref{FIG:original_signal}a, \ref{FIG:original_signal}b, 
\ref{FIG:VLD_waveform}c, \ref{FIG:SRA_12_waveform}e, and \ref{FIG:SRA_final_waveform}h represent the EEG epoch of different duration. The output of VLD contains unnecessary and unwanted pulses, as depicted on Fig. \ref{FIG:VLD_waveform}d. The SRA eliminates unwanted signals in every iteration. Figure \ref{FIG:SRA_12_waveform}f and 
\ref{FIG:SRA_12_waveform}g are the output of SRA after first iteration and second iteration. Figure \ref{FIG:SRA_final_waveform}i shows the initiation of seizure detection after $(n-k)$th iteration, which reports a lesser number of pulses on the non-seizure area. After $n$th iteration, SRA eliminates all the pulses on the non-seizure area and processes pulses on the ictal area and declares a seizure. 

Overall, the detector misses one seizure instance for the chosen EEG dataset. 
The sensitivity, and specificity of the seizure detector are measured as 96.9\%, and 97.5\%, respectively.The average latency of the proposed system is found as 3.6 seconds.The characterization of the seizure detector is shown in Table \ref{TABLE:characterization_seizure_detector}. 

   
\begin{table}[t]
	\centering
	\caption{\textcolor{black}{CHARACTERIZATION OF THE SEIZURE DETECTOR}}
	\label{TABLE:characterization_seizure_detector}
	\begin{tabular}{lc}
		\hline  \hline
		\addlinespace
		Parameter &  Value \\
		\hline \hline
		\addlinespace
		
		Seizure Frequency (Minimum) & 3 Hz  \\
		Seizure Frequency (Maximum) & 29 Hz  \\
		VLD (Average Lower Threshold) & 210 mV  \\
		VLD (Average Upper Threshold) & 380 mV  \\
		Sensitivity & 96.9\%\\
		Specificity & 97.5\%\\
		Power Consumption & 3.2 mW  \\
		\hline  \hline
	\end{tabular}
	
\end{table}

\begin{table}[t]
	\centering
	\caption{CHARACTERIZATION OF THE DRUG DELIVERY UNIT}
	\label{TABLE:characterization_drug_delivery_unit}
	\begin{tabular}{l c}
		 \toprule
		 \hline	
		 \addlinespace
		Parameters &  Value \\
		\hline \hline
		\addlinespace
		Piezoelectric (PZT) disc diameter & 9 mm\\
		Piezoelectric (PZT) disc thickness & 150 $\mu$m\\
		Membrane diameter & 10 mm\\
		Membrane thickness & 100 $\mu$m\\
		Young modulus (PDMS) & 0.8 MPa\\
		Possions ratio & 0.49\\
		Fluidic diodicity ($\eta$) & 2\\
		
		Liquid density & 1000 kg/m$^3$  \\
		
		
		\hline
		\hline
		
	\end{tabular}	
\end{table}

\begin{table}[t]
	\centering
	\caption{CHARACTERIZATION OF THE DRUG DELIVERY SYSTEM}
	\label{TABLE:characterization_drug_delivery_system}
\begin{tabular}{l c}
		 \hline \hline
		 \addlinespace
	
		Parameter &  Value \\
		\hline \hline
		\addlinespace
		Input voltage & 5 V\\
		Divergence angle (Diffuser) & 10$^\circ$ \\
		Frequency & 130 Hz  \\
		Power Consumption & 29.08 mW \\
		Volume flow & 3.08 ml/min  \\
		\hline
		\hline
	\end{tabular}
\end{table}

The divergence angle of the diffuser element was chosen as 10$^\circ$ which was found to be the optimal value. An increase in chamber diameter leads to a quadratic increase  in the net flow rate but also increases the dead volume which is undesirable for on-chip applications. Considering the dead volume and other factors, the chamber diameter was chosen as  10 mm. 

\begin{figure}[htbp]
	\centering
	\includegraphics[width=0.65\textwidth]{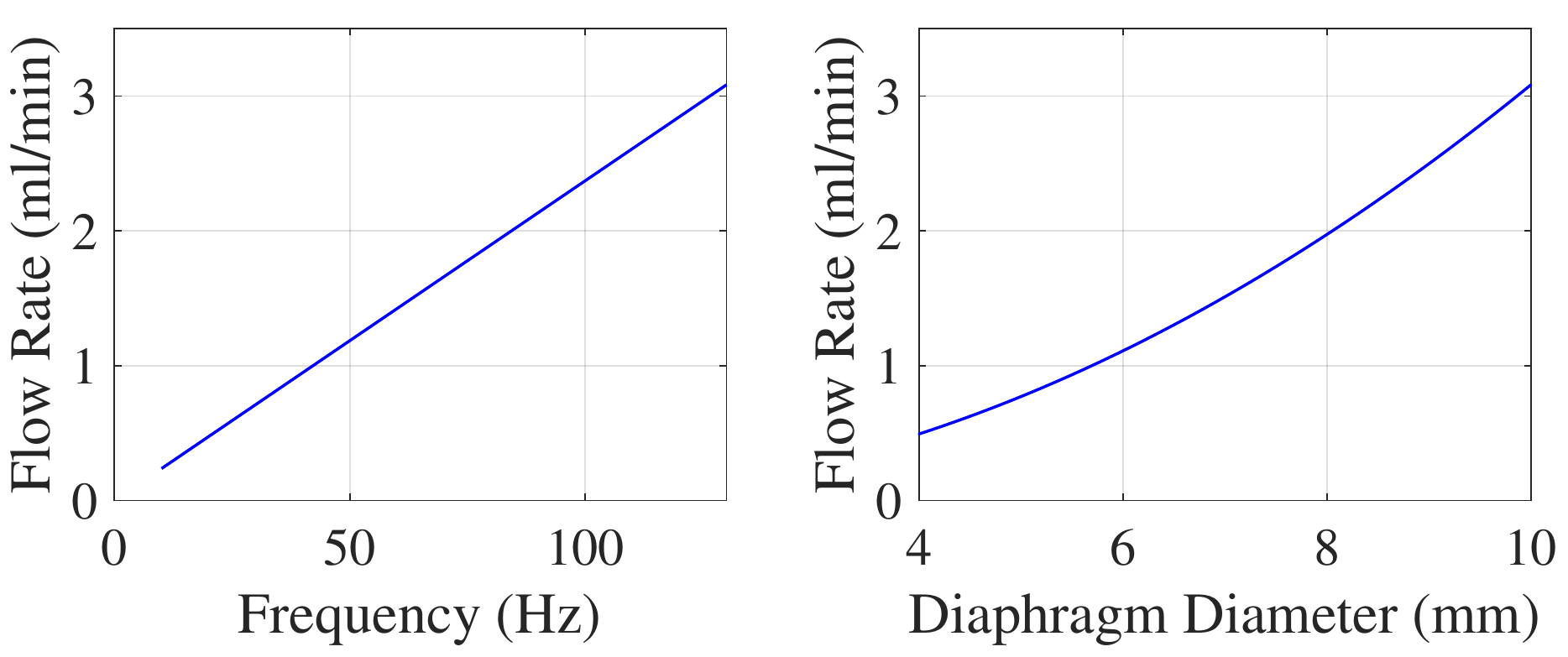}
	\caption{\textcolor{black}{Volumetric Flow Rate as a Function of (a) Actuation Frequency (b) Diaphragm Diameter.}}
	\label{FIG:frequency_diameterVSflow}
\end{figure}

The characterization of the drug delivery unit, is shown in Table \ref{TABLE:characterization_drug_delivery_unit}. Material with lower Young modulus has been chosen as a diaphragm membrane to enhance the volumetric flow rate. It is observed from the simulation data that an increase in diaphragm thickness leads to a reduction of volumetric flow rate. Simulation results also show that a small change in diaphragm diameter leads to a drastic change in the volumetric flow rate, while keeping diaphragm thickness constant at 100 $\mu$m. Fig.\ref{FIG:frequency_diameterVSflow}a shows the change in the flow rate with actuation frequency, while keeping diaphragm diameter at 10 mm and thickness at 100 $\mu$m .The variation of volumetric flow rate with diaphragm diameter is shown in Fig.\ref{FIG:frequency_diameterVSflow}b.  The drug delivery system has been characterized in Table \ref{TABLE:characterization_drug_delivery_system}. The proposed system provides a maximum flow rate of 3.08 ml/min. The total power consumption for the drug delivery system has been measured as 29.08 mW, which is significantly lower than to existing drug delivery systems. A comparison with  existing seizure detectors and drug delivery systems is provided in  Table \ref{TABLE:comparison_chart}.

	%

\begin{table*}[htbp]
	\centering
	\caption{\textcolor{black}{COMPARISON WITH EXISTING SYSTEMS}}
	\label{TABLE:comparison_chart}
	\begin{tabularx}{\textwidth}{@{} C C C C C C C C C @{}}
		\hline
		\hline
		\addlinespace
		Existing Works & Seizure Detection Method & Sensitivity & Specificity & Latency & Power Consumption on SD\textsuperscript{a} &  Drug Delivery Unit & Power Consumption on DDS\textsuperscript{b}\\
		\hline
		\hline
		\addlinespace
		
		Verma, et al. 2010 \cite{Verma_IEEE-JSSC_2010}   &   SVM\textsuperscript{c} &  \textgreater90\% &  NA & 5 sec & 120 $\mu$W & NA & NA \\
		\addlinespace  
		Salam, et al. 2012  \cite{Salam_IEEE-TBCS_2011}  &  AFED\textsuperscript{d} & 100 \% & 100 \% & 16 sec & 45.1 mW (AM\textsuperscript{e})  22.1 mW (PSM\textsuperscript{f}) & Micro-pump and HSE\textsuperscript{g} & 61.2 mW (AM\textsuperscript{e}) 33.1 mW (PSM\textsuperscript{f}) \\
		\addlinespace  
		
		Hamie, et al. 2013 \cite{Hamie_MeMeA_2013}  &  AFED\textsuperscript{b}  & NA & NA & NA & NA & Electromagnetic micro-pump & 40.2 mW  \\
		\addlinespace  
		
		Yoo, et al. 2013 \cite{Yoo_IEEE-JSSC_2013} &  LSVM\textsuperscript{h} &  84.4 \% & 96 \% & 2 sec & 1.49 $\mu$J/class (energy efficiency) & NA & NA \\
		\addlinespace  
		Altaf, et al. 2015 \cite{Altaf_IEEE-JSSC_2015} & LSVM\textsuperscript{h} & 95.7 \% & 98 \% & 1 sec & 2.73 $\mu$J/class (energy efficieny) & NA & NA \\
		\addlinespace 
		\textbf{Our Proposed System}  &   SRA\textsuperscript{i} & 96.9 \% & 97.5 \% & 3.6 sec  & 3.2 mW  & PZT based micro-pump & 29.08 mW  \\
		
			\bottomrule
		\bottomrule
		\addlinespace
	\end{tabularx}
	\caption*{\textsuperscript{a}SD: Seizure Detector,\textsuperscript{b}DDS: Drug delivery system, \textsuperscript{c}SVM: support vector machine, \textsuperscript{d}AFED : Asynchronous front-end detector,  \textsuperscript{e}AM: Active mode, \textsuperscript{f}PSM: Power saving mode,  \textsuperscript{g}HSE: Hybrid subdural electrode, \textsuperscript{h}LSVM: Linear support vector machine, \textsuperscript{i}SRA: Signal rejection algorithm}
\end{table*}


\section{Conclusions and Future Research}
\label{Sec:Conclusions}


We have proposed an automated seizure detector and drug delivery system for seizure control. The proposed system was implemented in a system level simulator. The system level simulation demonstrated the detection of seizure onset marked by hyper-synchronous activity. The rejection algorithm employed by the seizure detector is highly efficient in minimizing false detection by removing unwanted signals. The micro-pump system delivers drug into the onset area upon seizure detection and provides better control of refractory seizures. The simulation results indicate that there is a considerable reduction in power consumption compared to existing seizure detectors. In future research we will miniaturize the proposed system using the latest integrated circuit technologies for implantable drug delivery applications and test this solution in an animal model of epilepsy.


\section*{Acknowledgments}

This article is an extended version of the published article \cite{Sayeed_TCE_2019-Aug_eSeiz}. The current article has been thoroughly revised to add drug-delivery aspects which were not presented in \cite{Sayeed_TCE_2019-Aug_eSeiz}. The authors strongly believe that ``cSeiz'' of the current paper presents a unified perspective of seizure detection and control in one device instead of ``eSeiz'' of the published article, which provides a better perspective of this important problem of seizure and its control \cite{Sayeed_TCE_2019-Aug_eSeiz}.

The authors would like to acknowledge the inputs of Dr. Hitten P. Zaveri, Department of Neurology at Yale University, USA, on this work.



\bibliographystyle{IEEEtran}




\newpage
\section*{Authors' Biographies}

\begin{minipage}[htbp]{\columnwidth}

\vspace{1.0cm}
\begin{wrapfigure}{l}{1.40in}
	\vspace{-0.5cm}
\includegraphics[width=1.40in,keepaspectratio]{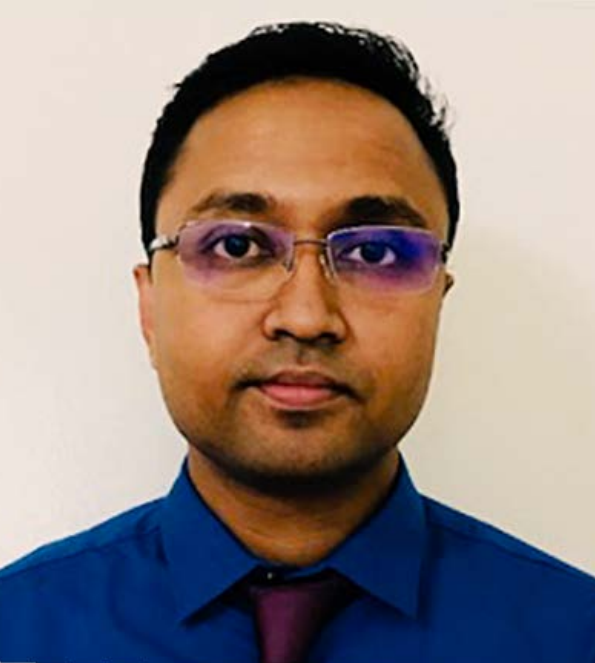}
	\vspace{-0.9cm}
\end{wrapfigure}
\noindent \textbf{Md Abu Sayeed} is currently a Ph.D. candidate in Computer Science and Engineering at the University of North Texas. He has received a Bachelors in electrical and electronic engineering from Khulna University of Engineering and Technology, Bangladesh, in 2009, and a Masters in Electrical Engineering from Lamar University, Beaumont, TX, in 2015. 
His research interest includes Biomedical Signal Processing for Smart Healthcare, Machine Learning, and Low-Power VLSI.
\end{minipage}\hfill
\vspace{2.5cm}
\begin{minipage}[thbp]{\columnwidth}
\begin{wrapfigure}{l}{1.40in}
	\vspace{-0.5cm}
	\includegraphics[width=1.40in,keepaspectratio]{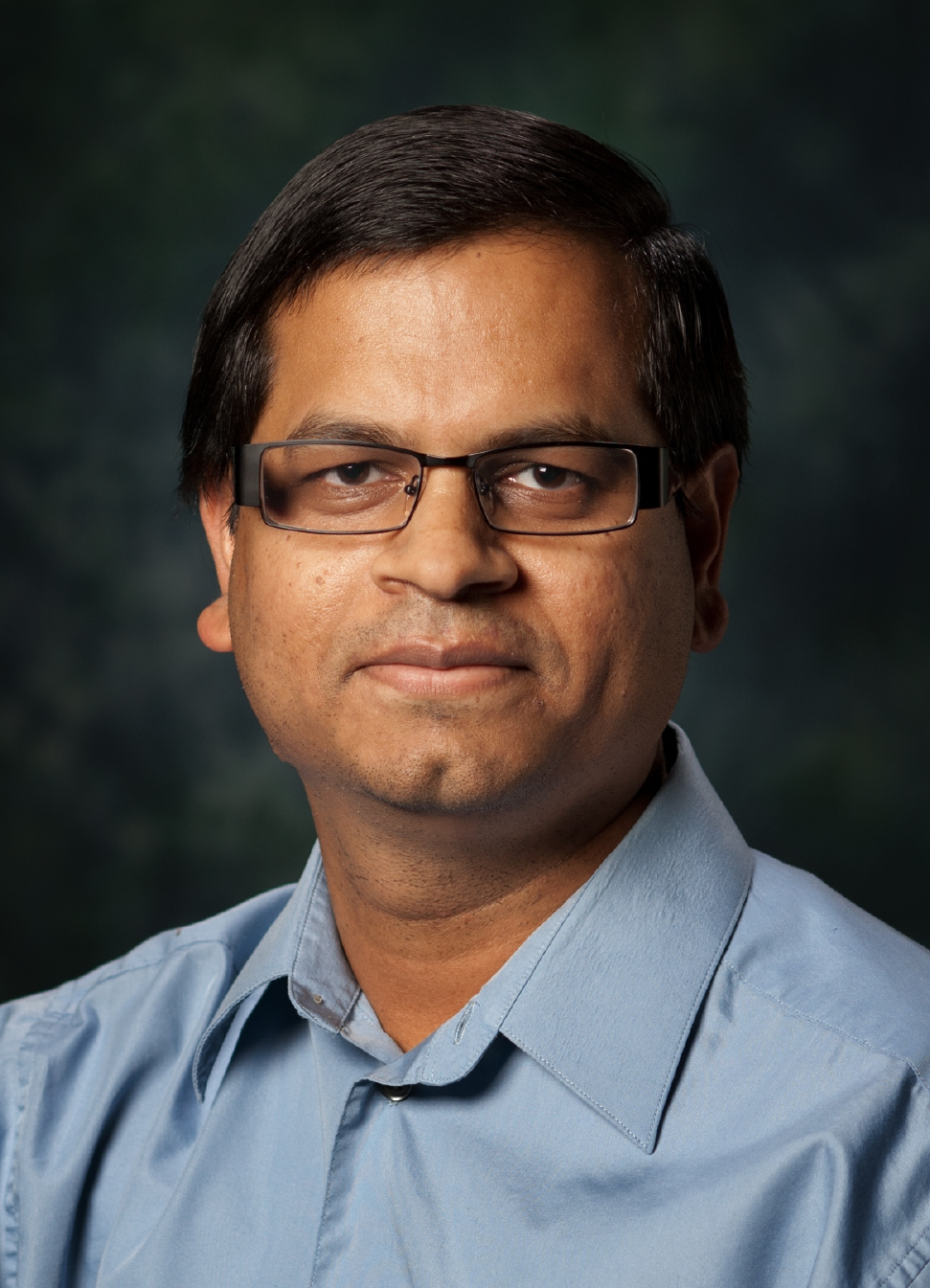}
	\vspace{-0.9cm}
\end{wrapfigure}
\noindent \textbf{Saraju P. Mohanty} received the bachelor's degree (Honors) in electrical engineering from the Orissa University of Agriculture and Technology, Bhubaneswar, in 1995, the master's degree in Systems Science and Automation from the Indian Institute of Science, Bengaluru, in 1999, and the Ph.D. degree in Computer Science and Engineering from the University of South Florida, Tampa, in 2003. He is a Professor with the University of North Texas. His research is in ``Smart Electronic Systems'' which has been funded by National Science Foundations (NSF), Semiconductor Research Corporation (SRC), U.S. Air Force, IUSSTF, and Mission Innovation Global Alliance. He has authored 300 research articles, 4 books, and invented 4 U.S. patents. His has Google Scholar citations with an H-index of 32 and i10-index of 110. He was a recipient of nine best paper awards, the IEEE-CS-TCVLSI Distinguished Leadership Award in 2018 for services to the IEEE and to the VLSI research community, and the 2016 PROSE Award for Best Textbook in Physical Sciences and Mathematics category from the Association of American Publishers for his Mixed-Signal System Design book published by McGraw-Hill. He has delivered 8 keynotes and served on 5 panels at various International Conferences.  He has been serving on the editorial board of several peer-reviewed international journals, including IEEE Transactions on Consumer Electronics (TCE), IEEE Transactions on Big Data (TBD), IEEE Transactions on Computer-Aided Design of Integrated Circuits and Systems (TCAD), and ACM Journal on Emerging Technologies in Computing Systems (JETC). He is currently the Editor-in-Chief (EiC) of the IEEE Consumer Electronics Magazine (MCE). He served as a founding Editor-in-Chief (EiC) of the VLSI Circuits and Systems Letter (VCAL) during 2015-2018. He has been serving on the Board of Governors (BoG) of the IEEE Consumer Electronics Society, and has served as the Chair of Technical Committee on Very Large Scale Integration (TCVLSI), IEEE Computer Society (IEEE-CS) during 2014-2018. He is the founding steering committee chair for the IEEE International Symposium on Smart Electronic Systems (iSES), steering committee vice-chair of the IEEE-CS Symposium on VLSI (ISVLSI), and steering committee vice-chair of the OITS International Conference on Information Technology (ICIT). He has mentored 2 post-doctoral researchers, and supervised 10 Ph.D. dissertations, 26 M.S. theses, and 10 undergraduate projects. 

\end{minipage}
\begin{minipage}[htbp]{0.99\columnwidth}
\vspace{1.0cm}

\begin{wrapfigure}{l}{1.4in}
	\vspace{-0.5cm}
	\includegraphics[width=1.4in,keepaspectratio]{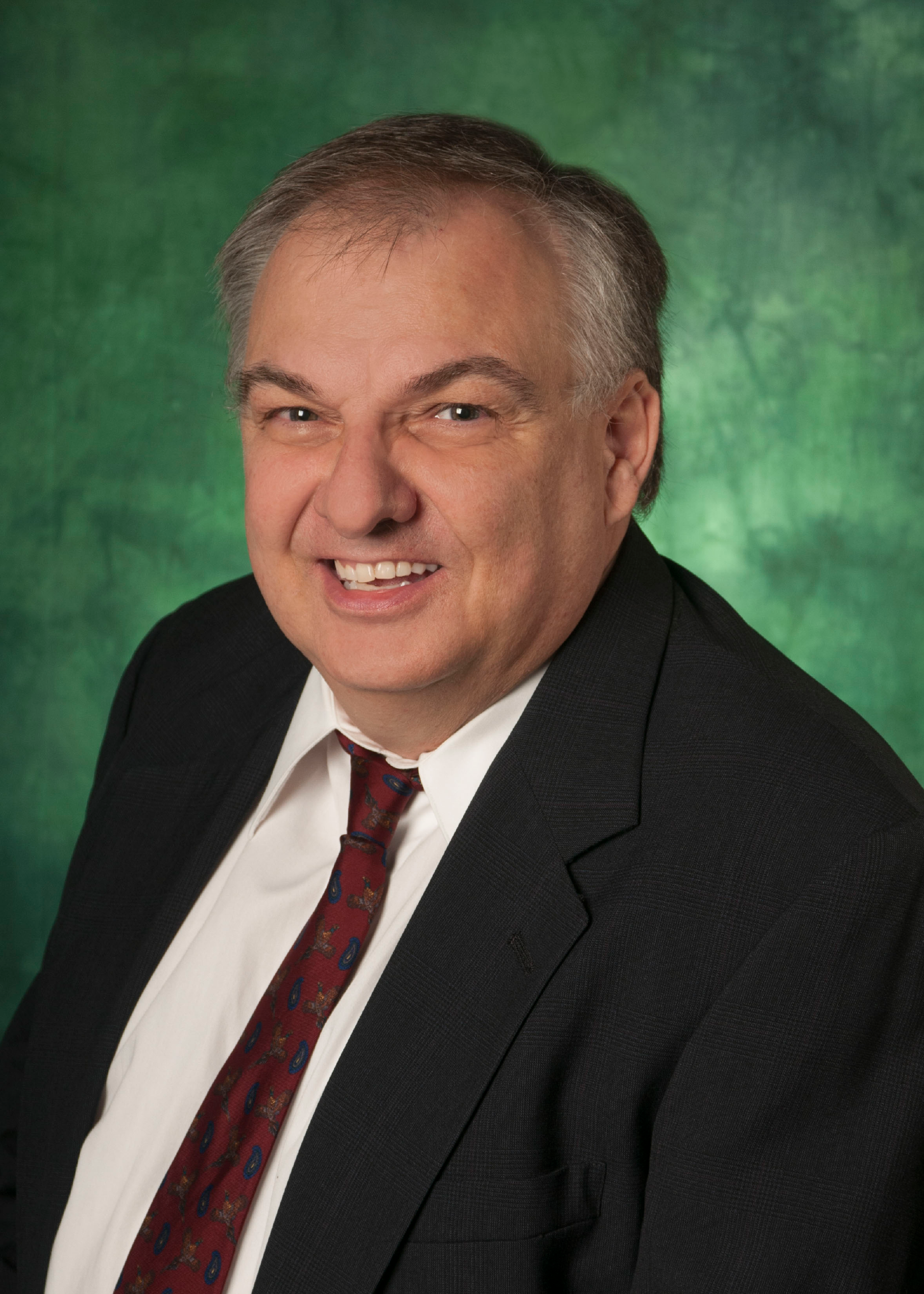}
	\vspace{-0.9cm}
\end{wrapfigure}
\noindent
\noindent \textbf{Elias Kougianos} received a BSEE from the University of Patras, Greece in 1985 and an MSEE in 1987, an MS in Physics in 1988 and a Ph.D. in EE in 1997, all from Lousiana State University.  From 1988 through 1997 he was with Texas Instruments, Inc., in Houston and Dallas, TX. Initially he concentrated on process integration of flash memories and later as a researcher in the areas of Technology CAD and VLSI CAD development. In 1997 he joined Avant! Corp. (now Synopsys) in Phoenix, AZ as a Senior Applications engineer and in 2001 he joined Cadence Design Systems, Inc., in Dallas, TX as a Senior Architect in Analog/Mixed-Signal Custom IC design. He has been at UNT since 2004. He is a Professor in the Department of Engineering Technology, at the University of North Texas (UNT), Denton, TX. His research interests are in the area of Analog/Mixed-Signal/RF IC design and simulation and in the development of VLSI architectures for multimedia applications. He is an author of over 120 peer-reviewed journal and conference publications.

\end{minipage}

\end{document}